\documentclass[osajnl,twocolumn,showpacs,superscriptaddress,10pt]{revtex4-1} %% use 11pt for Applied Optics
\usepackage{amsmath,amssymb,graphicx}

\begin{document}

\title{Propagation of fully-coherent and partially-coherent complex scalar fields in aberration space}

\author{David M. Paganin}
\affiliation{School of Physics and Astronomy, Monash University, Victoria 3800, Australia}

\author{Timothy C. Petersen}
\affiliation{School of Physics and Astronomy, Monash University, Victoria 3800, Australia}

\author{Mario A. Beltran}%
\affiliation{School of Science, RMIT University, Victoria 3001, Australia}
\affiliation{EMPA Swiss Federal Laboratories for Material Science and Technology, Switzerland}

\begin{abstract}

We consider the propagation of both fully coherent and partially coherent complex scalar fields, through linear shift-invariant imaging systems.  The state of such imaging systems is characterized by a countable infinity of aberration coefficients, the values for which can be viewed as coordinates for an infinity of orthogonal axes that span the ``aberration space'' into which the output propagates.  For fully coherent complex scalar disturbances, we study the propagation of the field through the imaging system, while for partially coherent disturbances it is the two-point correlation functions whose propagation we study.  For both systems we write down generalized propagators in both real and Fourier space, differential equations for evolution through aberration space, transport equations, and Hamilton--Jacobi equations. A generalized form of the Wolf equations is a special case of our formalism.  \end{abstract}

\maketitle  

\section{Introduction} \label{Intro}

Imaging systems are a formidable tool for studying the structure of matter. Applications are legion, ranging across the full spectrum from biomedical sciences to materials engineering, from telescopes to microscopes. Such applications utilize radiation wavefields such as visible-light and x-ray photons, and matter wavefields such as electrons, neutrons and cold atoms. The rich utility and multiplicity of imaging scenarios and modalities continues to expand its key influence in many scientific disciplines.

All imaging systems are by nature imperfect. Such imperfections may be spoken of as ``aberrations'', referring to the fact that the (typically two-dimensional) intensity of the output field is different from that of the field input into the imaging system, ignoring multiplicative and transverse scale factors. The concept of optical aberrations, in addition to its loose colloquial usage to denote distortions or imperfections in the output intensity map, has been rendered precise in both the geometric-optics \cite{BornWolf1999} and wave-optics \cite{CowleyBook} formalisms for imaging systems.  

For both the geometrical and the wave theories of imaging systems, various classes of aberration have attracted historical terms such as piston, tip, tilt, coma, astigmatism, defocus, spherical aberration, higher-order spherical aberration, Gaussian blur, trefoil, tetrafoil etc. \citep{BornWolf1999,AbbBalancingPaganinGureyev}. There is a close correspondence between the set of all possible optical aberrations, and the complete orthogonal set of functions over the two-dimensional disc known as the Zernike polynomials \cite{Lakshminarayanan}. In turn, the Zernike polynomials can be put into one-to-one correspondence with Cartesian polynomials in the transverse coordinates $x$ and $y$ \cite{Lakshminarayanan}.

Notwithstanding the negative implications associated with the term ``aberration'', in some cases the presence of aberrations may prove beneficial for an imaging system. For example, if a uniformly illuminated specimen is both non-absorbing and optically thin, then the intensity of the field over its exit surface will contain little to no contrast. For such an object, whose exit-surface wavefield corresponds to the entrance surface of an imaging system, a lack of aberrations in the system implies a corresponding lack of contrast in the associated image.  In such a case aberrations can be deliberately introduced to render visible the object features. Techniques such as Zernike phase-contrast optical microscopy \cite{Zernike}, inline holography \cite{Gabor1948} and Schlieren imaging \cite{Settles2001} exploit these imperfections. In such contexts the term ``aberration'' is something of a misnomer, since such aberrations improve rather than degrade the image, by rendering output intensity maps that do not solely depend on the near-featureless intensity distribution of the input maps, but also depend upon the input phase distribution due to the thin non-absorbing object. 

Another imperfection may be associated with the wavefield itself. Despite the great advancements in optical technology which have produced wavefields with a very high degree of spatial and temporal coherence, particularly in the visible-light optics regime since the advent of the optical laser, it is fundamentally impossible to reach perfect coherence. This limitation can be understood as a direct consequence of the optical uncertainty principle \cite{Mansuripur}, the time--bandwidth form of which implies that any optical signal which exists for a finite time must have a non-zero bandwidth of frequencies. The field therefore has non-perfect temporal coherence, and therefore cannot reach the limit of perfect coherence. A related but distinct constraint is furnished by the noise--resolution uncertainty principle \citep{Gureyev2016,GureyevKozlov}, which links the finite number of quanta (e.g. photons, electrons, neutrons etc.) passed through a given optical imaging system, to the maximum spatial resolution that is achievable by such a system.  

Taken together, the imperfections associated with both optical aberrations and partial coherence play an essential role in wave optics. The previously mentioned examples of Zernike phase contrast \cite{Zernike}, inline holography \cite{Gabor1948}  and Schlieren imaging \cite{Settles2001} remain relevant in this broader context. Another class of examples is given by various means for X-ray phase contrast imaging with low-coherence laboratory-based sources, utilizing aberrations to enhance image contrast \citep{Wilkins,PaganinAlg2002}. This is now becoming the norm and sees enormous potential in clinical applications \citep{Wilkins,Pavlov,Pfeiffer,Olivo}. In the domain of cold atom optics, magnetically trapped Bose--Einstein condensates (BEC) can evolve into specific caustic structures due to imperfect focusing as a consequence of induced atom-optical aberrations \cite{Simula}. In visible light microscopy,  techniques such as variable coherence tomography and depth-moments extraction use coherence variation to retrieve structural information of a specimen \citep{Baleine,BeltranMoments2015}. Quantitative optical phase microscopy \citep{PaganinNungent,BartyNungent} uses the defocus aberration to the same end, again using partially coherent light. These examples motivate a generalized theory which describes the evolution and energy transport of partially coherent optical wavefields of arbitrary aberrated imaging systems.

It is in this context that the present paper explores a generalized diffraction theory for aberrated imaging systems, via the concept of “aberration space” originally proposed by Allen {\em et al.} \cite{AllenOxleyPaganin}. This construct considers a given two-dimensional forward-propagating coherent complex scalar field that propagates through an infinite multiplicity of possible aberrated shift-invariant linear imaging systems.  The associated imaging-system control parameters, given by the set of all possible values of the aberration coefficients associated with the set of all possible aberrated linear imaging systems, is viewed as furnishing coordinates for an infinite dimensional aberration space, into which the input field propagates.  

The present paper generalizes this previous work by: (i) extending the investigation of Allen {\em et al.} \cite{AllenOxleyPaganin}, which was limited to coherent aberrations (i.e. unitary shift-invariant aberrated linear imaging systems) to include both coherent and incoherent aberrations; (ii) obtaining generalized wave, continuity and eikonal (Hamilton--Jacobi) equations for propagation in aberration space, as well as writing a generalized diffraction integral in both real-space and Fourier-space forms; (iii) extending the formalism to statistically stationary partially coherent complex scalar wavefields, as quantified by their cross-spectral density, for which generalized wave equations (generalized Wolf equations), continuity and eikonal equations are obtained.

The paper is structured as follows. Section II revises the aberration-coefficient formulation of the transfer function formalism for linear shift-invariant imaging systems. This involves a power series expansion of the complex argument of the Fourier representation of the propagator associated with such imaging systems. The corresponding coefficients in the power series are generalized aberration coefficients, special cases of which can be associated directly with classical aberrations such as defocus, astigmatism, spherical aberration, Gaussian damping etc. This last-mentioned connection is natural, in light of the previously mentioned close correspondence between classical aberrations and Zernike polynomials, coupled with the low-order Cartesian polynomial representation of the Zernike polynomials. Section III treats the propagation of coherent complex scalar fields into aberration space. A generalized Huygens-type construction is given for propagation into such space, both in convolution form and operator form. Generalized wave equations, continuity equations and eikonal equations are written, with several special cases being given to exemplify the broad scope of the formalism. Two distinct classes of aberration are identified: (i) those associated with the vacuum wave equation of the field under consideration (which may, as we show, obey an infinite multiplicity of vacuum wave equations, and therefore correspond to an infinite multiplicity of classical scalar field theories), and (ii) those aberrations associated with a linear shift invariant optical imaging system through which the said field is transmitted. Section IV generalizes all of these results to the case of partially coherent complex scalar fields in which a specified cross-spectral density (two-point correlation function in the space--frequency domain) propagates through an aberrated shift-invariant imaging system. This generalization is achieved via the space--frequency description of statistically stationary partially coherent fields \citep{MandelWolfBook,WolfSpaceFreqPaper,WolfIntroToCoherenceTheory}, a formalism that is closely related to the density-matrix formalism of quantum mechanics \cite{Messiah}. Here, the stochastic process representing the partially coherent field is obtained via a suitable average over a statistical ensemble of strictly monochromatic fields, all of which have the same angular frequency, for each pair of spatial coordinates and each angular frequency.  We discuss some of the broader implications of our results, including possible avenues for future work, in Sec. V.  

\section{Background} \label{Background}

\begin{figure}[h]
\centering
\includegraphics[scale=0.19]{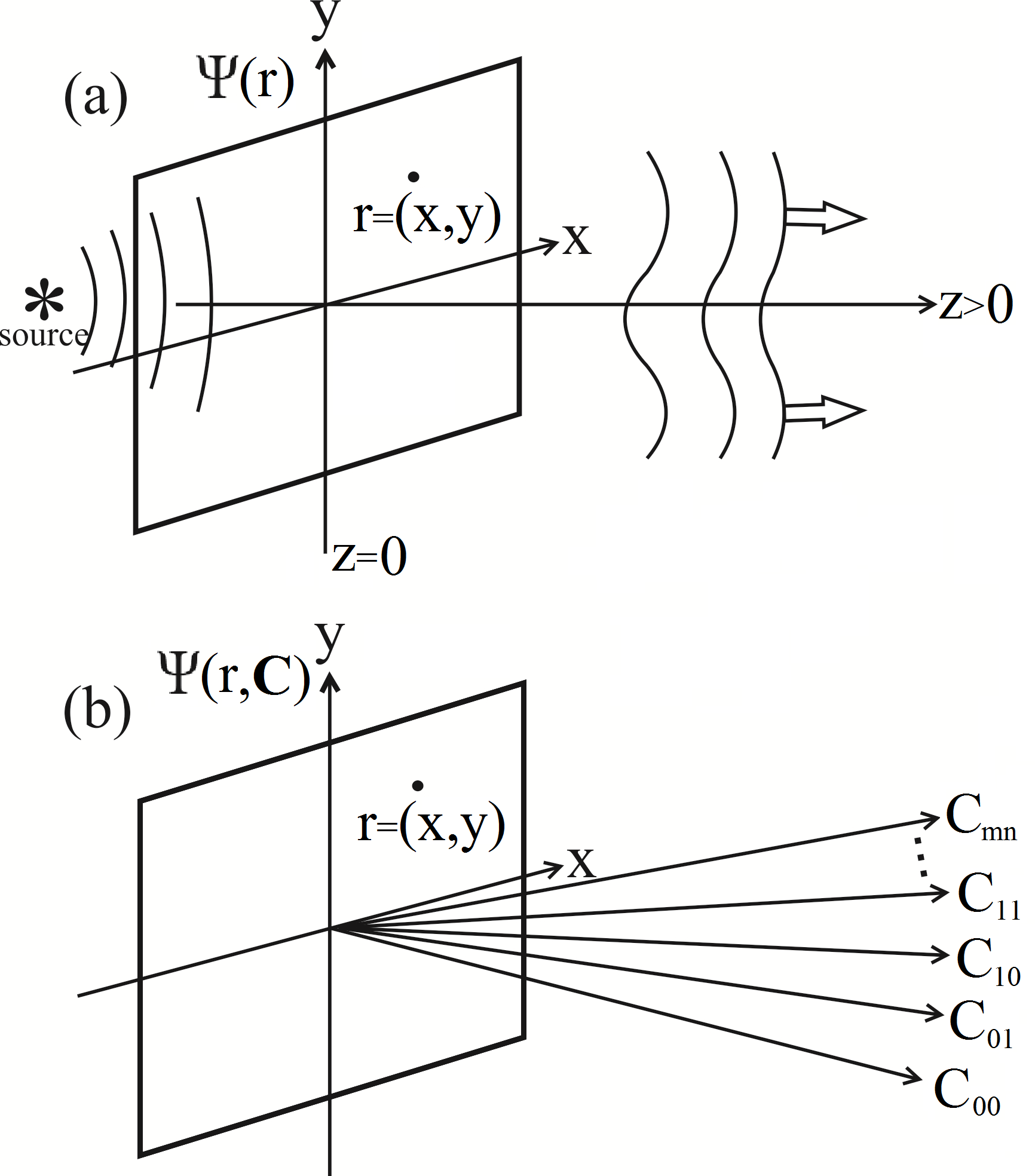}
\caption{(a) Illustration of different cases of wave propagation. (a) 2-dimensional fully coherent complex scalar paraxial waves propagating along the optic axis $z$. (b) 2-dimensional fully coherent complex scalar waves propagating in aberration space.}\
\label{figure1}
\end{figure}

With reference to Fig.~\ref{figure1} (a), recall the following form of the Fresnel diffraction integral, for evolving a forward-propagating paraxial monochromatic complex scalar plane wave from its boundary value $\Psi({\bf{r}} | z=0)$ over the plane $z=0$, to its boundary value over some downstream plane $z \ge 0$ \cite{NazarathyShamir}:

%%EQUATION------------------------
\begin{eqnarray} 
\Psi \left ( \textbf{r}\mid z\geq  0 \right ) =\frac{1}{2\pi}\iint_{-\infty}^{\infty}d\textbf{k}_{\textbf{r}}\widehat{\Psi}( \textbf{k}_{\textbf{r}}\mid z=0) \nonumber\\
 \times \exp \left [-i\frac{z\left | \textbf{k}_{\textbf{r}} \right |^{2}}{2k} +i\textbf{k}_{\textbf{r}}\cdot \textbf{r}\right ]. \nonumber\\
\label{AWECSD00}
\end{eqnarray}
%---------------------------------

\noindent Here,  $\textbf{r}=(x,y)$ denotes the position vector in the plane perpendicular to the optic axis $z$, a caret denotes Fourier transformation with respect to $x$ and $y$, $\textbf{k}_{\bf r} = (k_x,k_y)$ denotes the Fourier (spatial frequency) coordinates dual to $(x,y)$, and $k = 2\pi/\lambda$ is the wavenumber corresponding to the wavelength $\lambda$.

%%%%%%%
%FIGURE
%%%%%%%%%%%%%%%%%%%%%%%%%%%%%%%%%%%%%%%%%%%%%%%%%%%%%%%%%%%%%%%%%%%%%%
%%%%%%%%%%%%%%%%%%%%%%%%%%%%%%%%%%%%%%%%%%%%%%%%%%%%%%%%%%%%%%%%%%%%%%

The forward and inverse Fourier transform conventions used here are: 

%%EQUATION------------------------
\begin{subequations}
\begin{align}
\widehat{\Psi}( \textbf{k}_{\textbf{r}}) =\frac{1}{2\pi}\iint_{-\infty}^{\infty}d\textbf{r}\exp(-i\textbf{k}_{\textbf{r}}\cdot \textbf{r})\Psi(\textbf{r}) \label{AWECSD02:a}, \\
 \Psi(\textbf{r}) =\frac{1}{2\pi}\iint_{-\infty}^{\infty}d\textbf{k}_{\textbf{r}}\exp(i\textbf{k}_{\textbf{r}}\cdot \textbf{r})\widehat{\Psi}( \textbf{k}_{\textbf{r}}).
 \label{AWECSD02:b}
\end{align}
\end{subequations}
%---------------------------------

Equation~\ref{AWECSD00} can be readily generalized, by recognizing that the Fourier-space representation of the Fresnel propagator (corresponding to the quadratic term in square brackets) can be replaced by the Fourier-space representation of a much more general propagator \cite{GoodmanFourierOpticsBook,AllenOxleyPaganin,PaganinXRayBook,AbbBalancingPaganinGureyev}.  This more general propagator may be associated with changing the differential equation obeyed by the complex scalar field {\em {in vacuo}}, and/or interposing a shift-invariant linear imaging system \cite{GoodmanFourierOpticsBook} in between the entrance surface $z=0$ and the exit surface of the system.  

Assuming the Fourier representation of the said generalized propagator to be sufficiently well behaved that its logarithm can be expanded as a Taylor series in $(k_x,k_y)$ \cite{AllenOxleyPaganin,AbbBalancingPaganinGureyev}, one can generalize Eq.~\ref{AWECSD00} by replacing the quadratic term in square brackets with an arbitrary Taylor series with complex coefficients $C_{mn}$.  This gives \cite{AllenOxleyPaganin,PaganinXRayBook,AbbBalancingPaganinGureyev}: 

%%EQUATION------------------------
\begin{eqnarray} 
\Psi \left ( \textbf{r}\mid \left \{ C_{mn} \right \} \right ) =\frac{1}{2\pi}\iint_{-\infty}^{\infty}d\textbf{k}_{\textbf{r}}\widehat{\Psi}( \textbf{k}_{\textbf{r}}\mid \left \{ C_{mn}\right \}=0) \nonumber\\
 \times \exp \left [ i\sum_{m,n}C_{mn}k^{m}_{x}k^{n}_{y}+i\textbf{k}_{\textbf{r}}\cdot \textbf{r}\right ]. \nonumber\\
\label{AWECSDa0}
\end{eqnarray}
%---------------------------------

\noindent Here, $\Psi \left ( \textbf{r}\mid \left \{ C_{mn} \right \} \right )$ is the aberrated complex scalar wave-field, $m,n=0,1,2,...$, $\widehat{\Psi}( \textbf{k}_{\textbf{r}}\mid \left \{ C_{mn}\right \}=0)$ is the Fourier transform of $\Psi \left ( \textbf{r}\mid \left \{ C_{mn} \right \}=0 \right )$ with respect to $(x,y)$ and the shorthand $ \left \{ C_{mn} \right \}=0 $ indicates that $\{ C_{mn} \} = \{0, 0, 0, \cdots \}$.

Adopt the language of physical optics and denote as {\em aberration coefficients} \cite{PaganinXRayBook,AbbBalancingPaganinGureyev} the set of coefficients $\{ C_{mn} \}$.  The coefficients $\{ C_{mn}\}$ completely characterize the state of the imaging system.  Each of the $C_{mn}$ coefficients is complex, and can be written in the form

%%EQUATION------------------------
\begin{eqnarray} 
C_{mn} \equiv C_{mn}^{(R)}+iC_{mn}^{(I)},
\end{eqnarray}
%---------------------------------

\noindent where $C_{mn}^{(R)}$ and $C_{mn}^{(I)}$ are real numbers that denote the real and imaginary parts of $C_{mn}$, respectively.  The set $\{C_{mn}^{(R)}\}$ denotes the so-called coherent aberrations, with $\{C_{mn}^{(I)}\}$ being the set of incoherent aberrations.

The reasonable physical assumptions that 

%%EQUATION------------------------
\begin{eqnarray} 
\nonumber C_{mn}^{(I)} \ge 0 \quad {\textrm{for all}}~\it{m}~\textrm{and}~{\it n}, \\ C_{mn}^{(I)} = 0 \quad {\textrm{if}}~m~\textrm{and/or}~n~{\textrm{is odd}}, 
\label{AbbSpaceIsAHalfSpace}
\end{eqnarray}
%---------------------------------

\noindent amount to assuming that incoherent aberrations may suppress the Fourier coefficients of the input field, but may not amplify them.  This is consistent with the assumption that no gain media be present in the linear shift-invariant system.  Stated differently, we assume that energy associated with any Fourier degree of freedom may be blocked/damped upon passage from the entrance to the exit of the linear imaging system, but never amplified.  

We give some examples to relate the set $\{ C_{mn} \}$ to the forms for the aberrations (e.g. defocus, spherical aberration, Gaussian blurring) that are often encountered in optical imaging theory. (a) If we choose

%%EQUATION------------------------
\begin{eqnarray} 
 C_{20}=C_{02}= -\frac{z}{2k} 
\label{DP300}
\end{eqnarray}
%---------------------------------
 
\noindent as the only non-zero members of the set of aberration coefficients $\{ C_{mn} \}$, Eq.~(2) reduces to the form of the Fresnel diffraction integral given in Eq. (1). Thus $C_{20}$ and $C_{02}$ are defocus aberrations, in the $x$ and $y$ directions respectively. (b) If we choose $C_{mn}$ such that \cite{AbbBalancingPaganinGureyev}

%%EQUATION------------------------
\begin{eqnarray} 
C_{04}=C_{40}=\frac{1}{2}C_{22}=\Xi \: \in \mathbb{R}, \: \Xi=\frac{C_{S} }{8 k^{3}}  
\label{DP301}
\end{eqnarray}
%---------------------------------

\noindent are the only non-zero coefficients, then $\sum_{m,n}C_{mn}k^{m}_{x}k^{n}_{y}$ reduces to $\Xi (k^{2}_{x}+k^{2}_{y})^{2}$ and $\Xi$ is proportional to the spherical aberration $C_{S}$ in the shift-invariant linear imaging system. (c) If we choose the only non-zero coefficients to be such that $\exp ( i\sum_{m,n}C_{mn}k^{m}_{x}k^{n}_{y})$ reduces to $\exp [ -\Theta ( k^{2}_{x}+k^{2}_{y})]$, where $\Theta$ is real and non-negative, then one has Gaussian blurring at the level of the complex field.  Note that, in this case, all non-zero aberration coefficients are purely imaginary. Indeed, a necessary condition for damping of Fourier-space degrees of freedom, is that at least one of the aberration coefficients have a non-zero value.  (d) Other aberrations such as astigmatism, higher-order spherical aberration, coma, etc. can be readily incorporated, with all aberrations present being cascaded together \cite{AbbBalancingPaganinGureyev}. 

We close this section by noting that Eq.~\ref{AWECSDa0} has the real-space form given by the linear integral transform

%%EQUATION------------------------
\begin{eqnarray} 
\Psi \left ( \textbf{r}\mid \left \{ C_{mn} \right \} \right ) 
= \Psi \left ( \textbf{r}\mid \left \{ C_{mn} \right \} = 0\right) \circledast_2 K({\bf r}, \left \{ C_{mn} \right \}), \quad
\label{DP301aaa}
\end{eqnarray}
%---------------------------------

\noindent where $K({\bf r}, \left \{ C_{mn} \right \})$ is a real-space propagator and ``$\circledast_2$'' denotes two-dimensional convolution over ${\bf r}$.  The convolution theorem of Fourier analysis implies the kernel $K({\bf r}, \left \{ C_{mn} \right \})$, which may be viewed as a generalized Huygens-type wavelet, to be proportional to the inverse Fourier transform of $\exp(i\sum_{m,n}C_{mn}k_x^mk_y^n)$ with respect to ${\bf k}_{\bf r}$.  Recursive application of the Fourier convolution theorem, to the result in the previous sentence, implies that the propagator may be written as the following cascade of convolutions:

%%EQUATION------------------------
\begin{eqnarray} 
K({\bf r}, \left \{ C_{mn} \right \}) =  \left[\prod_{m,n} K({\bf r} ,   C_{mn} ) 
\circledast_2 \right] \delta({\bf r}), \label{DMP00}
\end{eqnarray}
%---------------------------------

\noindent where $K({\bf r}, C_{mn} )$ is the inverse Fourier transform of $2\pi \exp(iC_{mn}k_x^m k_y^n)$ with respect to ${\bf k}_{\bf r}$, and $\delta({\bf r})$ denotes the Dirac delta.  Note that the Dirac delta is there to ``clean up'' the final convolution symbol which will ``dangle'' at the end of the expression once all square-bracket terms are written out in full.  If one omits the Dirac delta from the right side of the above equation, then the resulting modified right side will correctly describe the diffraction {\em operator} (cf. Nazarathy and Shamir \cite{NazarathyShamir}), which acts from right to left on the field input into the two-dimensional linear shift invariant imaging system, to yield the associated output. 

\section{Propagation of coherent fields in aberration space} \label{Sec00}

The propagator in Eq.~\ref{AWECSDa0} is very general and applies to a far greater variety of fields than just paraxial monochromatic scalar electromagnetic fields propagating either {\em in vacuo} or through a linear shift-invariant imaging system.  Indeed, Eq.~\ref{AWECSDa0} applies to a rich variety of forward-propagating complex fields, both paraxial and non-paraxial, such as those governed by the Schr$\ddot{\textup{o}}$dinger equation, the Helmholtz equation or the Klein--Gordon equation.  Again, Eq.~\ref{AWECSDa0} can apply to this rich variety of forward-propagating fields, whether they propagate {\em in vacuo} or through a shift-invariant linear imaging system.

These points are developed in the present section. In view of its length, it is broken up into several sub-sections. Sub-Section \ref{SubSec00A} introduces the concept of an infinite-dimensional aberration space \cite{AllenOxleyPaganin} with coordinates given by the set of aberration coefficients $\{C_{mn}\}$, together with the transverse coordinates $(x,y)$ of both the input and output imaging planes of the associated linear shift-invariant imaging system.  Sub-Section \ref{SubSec00B} derives wave, continuity and Hamilton--Jacobi (eikonal) equations for coherent forward-travelling scalar fields as they evolve along a particular direction in this aberration space. Some special cases of the aberration-space generalized wave and continuity equations are given in Subsec. \ref{SubSec00C}. Sub-section \ref{SubSec00D} gives a factorization of the aberration propagator, to separate out the effects of the particular field equation that a given forward propagating complex scalar field obeys {\em in vacuo}, from the action upon that field of a linear shift-invariant optical imaging system.  

\subsection{The concept of aberration space} \label{SubSec00A}

Returning attention to Eq.~\ref{AWECSDa0}, we have already stated that it can be viewed as describing the propagation of the boundary value $\Psi(\textbf{r}|z=0)$ of our forward propagating scalar complex field, to a two-spatial-dimensional output plane $(x,y)$.  The equation therefore describes propagation into the ``physical space'' $z \ge 0$, with the control parameter describing the state of the shift-invariant system being a vector $\bf C$ in the infinite-dimensional control space whose components are given by any particular choice of aberration coefficients $\{C_{mn}\}$.  

Alternatively, one can view the totality of all possible output fields corresponding to a specified input field, for every possible state of the imaging system, as corresponding to the propagation of the input two-dimensional field into an infinite dimensional ``aberration space'' coordinatized by the components of $\bf C$ -- see Fig.~\ref{figure1} (b).  Under this view, each complex aberration coefficient $C_{mn}$ gives two real axes in the aberration space, corresponding respectively to a coherent aberration $C_{mn}^{(R)}$ and an incoherent aberration $C_{mn}^{(I)}$.  The vector $\bf C$ is therefore considered to have {\em real} components.  Coordinates along axes corresponding to the real part $C_{mn}^{(R)}$ of each $C_{mn}$ may take any real value, but Eq.~(\ref{AbbSpaceIsAHalfSpace}) implies that the corresponding imaginary parts $ C_{mn}^{(I)}$ are restricted to have $C_{mn}^{(I)} \ge 0$ when $m$ and $n$ are both even, and restricted to $C_{mn}^{(I)} = 0$ otherwise (i.e. when either or both of $m$ and $n$ are odd).

Coherent aberrations preserve the norm of the wavefunction, whereas incoherent aberrations do not.  If the only non-zero aberration coefficients belong to the set $\{C_{mn}^{(R)}\}$, then the resulting generalized diffraction operator $D(\{C_{mn}^{(R)}\})$,  mapping input to output field via 

%%EQUATION------------------------
\begin{eqnarray} 
\Psi \left ( \textbf{r}\mid \left \{ C_{mn}^{(R)} \right \} \right ) 
= D(\{C_{mn}^{(R)}\}) \Psi \left ( \textbf{r}\mid \left \{ C_{mn} \right \} = 0\right), \nonumber\\ 
\label{DMP-3}
\end{eqnarray}
%---------------------------------

\noindent is unitary and therefore both (i) energy-conserving and (ii) information-preserving in a formal sense. Both unitary and non-unitary forms of the operator $D(\{C_{mn}\})$ may be written in a Fourier representation as \cite{AllenOxleyPaganin}

%%EQUATION------------------------
\begin{eqnarray} 
D(\{C_{mn}\})=\mathcal{F}^{-1}\exp\left(i\sum_{m,n}C_{mn}k_x^mk_y^n\right) \mathcal{F}, 
\label{DMP-1}
\end{eqnarray}
%---------------------------------

\noindent where $\mathcal{F}$ denotes Fourier transformation with respect to $x$ and $y$, $\mathcal{F}^{-1}$ denotes the corresponding inverse transformation, and all operators act from right to left.  The operator $D(\{C_{mn}\})$ ceases to be unitary if any of the aberration coefficients have a non-zero imaginary component.  Equation Eq.~\ref{DMP-1} may be written in the cascaded form:

%%EQUATION------------------------
\begin{eqnarray} 
D(\{C_{mn}\})=\prod_{m,n} \mathcal{F}^{-1}\exp\left(iC_{mn}^{(R)}k_x^mk_y^n\right) \mathcal{F} \nonumber \\ \times \prod_{m,n} \mathcal{F}^{-1}\exp\left(-C_{mn}^{(I)}k_x^mk_y^n\right) \mathcal{F}. 
\label{DMP-2}
\end{eqnarray}
%---------------------------------

\noindent Note that the condition given by Eq.~\ref{AbbSpaceIsAHalfSpace} is considered to be implicit in the above expression.

Equations~\ref{DMP-3}--\ref{DMP-2} give the means for propagating a given wavefield $\Psi({\bf r},{\bf C}={\bf 0})$ into the aberration space with coordinates given by $({\bf r},{\bf C})$.  One may also write down a real-space form of $D(\{C_{mn}\})$, which is equal to the right side of Eq.~(\ref{DMP00}), with the Dirac delta omitted.

\subsection{Wave, continuity and Hamilton--Jacobi (eikonal) equations for propagation in aberration space} \label{SubSec00B}

If one applies the differential operator $\partial / \partial {C_{m'n'}^{(R)}}$ to both sides of Eq.~\ref{AWECSDa0}, corresponding to a particular value $(m',n')$ of $(m,n)$, then upon dropping the primes one obtains the set of propagating-field equations

%%EQUATION------------------------
\begin{eqnarray} 
\left ( \frac{\partial}{\partial {C_{mn}^{(R)}}}-\frac{i}{i^{m+n}} \frac{\partial^{m}}{\partial x^{m}}\frac{\partial^{n}}{\partial y^{n}} \right )\Psi (\textbf{r}, {\bf C} )=0,
\label{DMP01}
\end{eqnarray} 
%---------------------------------

\noindent where ${\bf C} \equiv \{C_{mn}\}$.  Conversely, if one applies the differential operator $\partial / \partial {C_{m'n'}^{(I)}}$ to both sides of Eq.~\ref{AWECSDa0}, again corresponding to a particular value $(m',n')$ of $(m,n)$, one obtains the set of diffusive field equations

%%EQUATION------------------------
\begin{eqnarray} 
\left ( \frac{\partial}{\partial {C_{mn}^{(I)}}}+\frac{1}{i^{m+n}} \frac{\partial^{m}}{\partial x^{m}}\frac{\partial^{n}}{\partial y^{n}} \right )\Psi (\textbf{r}, {\bf C} )=0.
\label{DMP02}
\end{eqnarray}
%---------------------------------

Taken together, Eqs.~\ref{DMP01} and \ref{DMP02} are the desired differential equations governing the propagation of coherent complex scalar waves in our infinite dimensional aberration space.  For every possible value of $(m,n)$, this gives an infinite system of independent linear differential equations, each of which are first order with respect to the corresponding aberration coefficients ${C_{mn}^{(R)}}$ and ${C_{mn}^{(I)}}$.  

A convenient reformulation of the above system of equations (Eqs.~\ref{DMP01} and \ref{DMP02}) is motivated by considering a shift-invariant linear imaging system whose state evolves as a continuous differentiable function of a given real parameter $\tau$, so that for each $m$ and $n$ we may write:

%%EQUATION------------------------
\begin{eqnarray} 
C_{mn}(\tau)=C_{mn}^{(R)}(\tau)+ i C_{mn}^{(I)}(\tau).
\label{DMP07}
\end{eqnarray}
%---------------------------------

\noindent This indicates that every aberration coefficient depends continuously on the real parameter $\tau\geqslant 0 $. 
With minimal loss of generality, we assume that each $C_{mn}(\tau)$ has a finite first derivative with respect to $\tau$.  The chain rule then gives the following expression for the $\tau$-rate-of-change of the output wavefunction $\Psi({\bf r},{\bf C}(\tau))$:

%%EQUATION------------------------
\begin{eqnarray} 
\frac{\partial}{\partial\tau}\Psi({\bf r},{\bf C}(\tau)) \quad\quad\quad\quad\quad\quad\quad\quad\quad\quad\quad\quad\quad\quad\quad\quad\quad\quad\quad 
\label{DMP08}
\\ =\sum_{m,n} \left[ 
\frac{\partial C_{mn}^{(R)}(\tau)}{\partial\tau}\frac{\partial \Psi({\bf r},{\bf C}(\tau))}{\partial C_{mn}^{(R)}}
+\frac{\partial C_{mn}^{(I)}(\tau)}{\partial\tau}\frac{\partial\Psi({\bf r},{\bf C}(\tau))}{\partial C_{mn}^{(I)}}
\right]. \nonumber
\end{eqnarray}
%---------------------------------

\noindent Using Eqs.~\ref{DMP01} and \ref{DMP02}, for $\partial \Psi({\bf r},{\bf C}(\tau)
) / \partial C_{mn}^{(R)}$ and $\partial \Psi({\bf r},{\bf C}(\tau)
) / \partial C_{mn}^{(I)}$ respectively, transforms Eq.~\ref{DMP08} into the following
generalization of Eq. 7 in Allen et al. \cite{AllenOxleyPaganin}:

%%EQUATION------------------------
\begin{eqnarray} 
\frac{\partial}{\partial\tau} \Psi (\textbf{r}, {\bf C} (\tau)
)={\mathcal L} (\tau)\Psi (\textbf{r}, {\bf C} (\tau)
),
\label{DMP03}
\end{eqnarray} 
%---------------------------------

\noindent where ${\mathcal L} (\tau)$ is the following linear differential operator:

%%EQUATION------------------------
\begin{eqnarray}
\nonumber \mathcal{L}(\tau)=\quad\quad\quad\quad\quad\quad\quad\quad\quad\quad\quad\quad\quad\quad\quad\quad\quad  \\ 
\sum_{m,n} \frac{1}{i^{m+n}}\left[ 
i\frac{\partial C_{mn}^{(R)}(\tau)}{\partial\tau}
- \frac{\partial C_{mn}^{(I)}(\tau)}{\partial\tau}\right]\frac{\partial^m}{\partial x^m}\frac{\partial^n}{\partial y^n}.
\label{DMP09}
\end{eqnarray}
%---------------------------------

Leaving all aberration coefficients arbitrary in the above equations corresponds to the propagation of the boundary value $\Psi(\textbf{r}| \tau=0)$ through the totality of all possible linear shift-invariant imaging systems.  Particular cases of imaging system (and/or field) give a useful simplification which is explored further below.  

While we have denoted arbitrary states of the system with the vector ${\bf C}\equiv\{C_{mn}\}$ in the aberration space, corresponding to any choice for the set $\{C_{mn}\}$ of all aberration coefficients, we will use the vector ${\bf D}$ in the same aberration space to denote particular special cases for the state of this system.  We will consistently use the notation:  

%%EQUATION------------------------
\begin{eqnarray} 
\textbf{D}=\left \{ (C^{(R)}_{00}, C^{(I)}_{00}); (C^{(R)}_{01}, C^{(I)}_{01}); (C^{(R)}_{10}, C^{(I)}_{10});... \right \} 
\label{AWECSDaa3}
\end{eqnarray}
%---------------------------------

\noindent when both coherent and incoherent aberrations are present.  If only coherent aberrations are non-zero, we use the notation:

%%EQUATION------------------------
\begin{eqnarray} 
\textbf{D}=\left \{ C^{(R)}_{00}, C^{(R)}_{01}, C^{(R)}_{10}, C^{(R)}_{02}, C^{(R)}_{11}, C^{(R)}_{20},... \right \}. 
\label{DMP11}
\end{eqnarray}
%---------------------------------

\noindent We will occasionally refer to ${\bf D}$ as corresponding to a particular ``direction'' in the aberration control space.  Note that, for aberrations of a given order (i.e. a given fixed $m+n$) in either of the above vectors, the ordering is given in ascending numerical order of $10m+n$.

(a) We return to the special case of Fresnel diffraction, already considered above, to indicate the idea of propagation along a particular direction in aberration space.  To this end, recall that Fresnel diffraction (of paraxial monochromatic complex scalar waves {\em in vacuo}) corresponds to setting $C_{20}=C_{02}=-z/2k$.  In the present context, this corresponds to choosing the particular aberration-space direction (cf. Eq.~\ref{DMP11})

%%EQUATION------------------------
\begin{eqnarray} 
\textbf{F}=\left \{  0, 0, 0,- \frac{1}{2k}, 0, - \frac{1}{2k}, 0,... \right \}
\label{AWECSDaa0}
\end{eqnarray}
%---------------------------------

\noindent and then evolving the field along the set of points $\tau \textbf{F}$, $\tau=z\geqslant 0$ along the direction $\bf{F}$ in aberration space.  This corresponds to Fresnel propagation through the distance $z \geqslant 0$, with $z$ considered as an independent variable.  The well-known paraxial/parabolic equation, obeyed by the field in this case, follows directly from Eq.~\ref{DMP03} \cite{SalehTeich}.  Equation \ref{DP301aaa} becomes the convolution form of the Fresnel diffraction integral \cite{WinthropWorthington}, Eq.~\ref{DMP00} becomes the real-space form of the Fresnel propagator, and Eqs.~\ref{DMP-3}--\ref{DMP-2} correspond to the operator form for Fresnel diffraction \cite{NazarathyShamir}. 

(b) As a second special case, application of spherical aberration corresponds to choosing the particular aberration--space direction

%%EQUATION------------------------
\begin{eqnarray} 
&& C_{S}\textbf{S}= \nonumber\\
&& \left \{  0_{10}, C^{(R)}_{04}=\frac{-C_{S}}{8k^{3}},0,C^{(R)}_{22}=\frac{-C_{S}}{4k^{3}},0,C^{(R)}_{40}=\frac{-C_{S}}{8k^{3}},... \right \} \nonumber\\
\label{AWECSDaa2}
\end{eqnarray}
%---------------------------------

\noindent where $0_{10}$ denotes a sequence of ten zeros, and then evolving the field along the set of points $ C_{S} \textbf{S} $ in aberration space.  This corresponds to viewing $C_{S}$ as an independent variable that characterises an optical imaging system with pure spherical aberration.

(c) Our third special case is Gaussian blurring.  Recall our earlier statement that, if we choose the only non-zero aberration coefficients such that $\exp ( i\sum_{m,n}C_{mn}k^{m}_{x}k^{n}_{y})$ reduces to $\exp [ -\Theta ( k^{2}_{x}+k^{2}_{y})]$, where $\Theta\ge 0$, then one has Gaussian blurring at the level of the complex field. The corresponding non-zero aberration coefficients $C_{20}^{(I)}=C_{02}^{(I)}=\tau=\Theta$ give the aberration-space direction

%%EQUATION------------------------
\begin{eqnarray} 
{\bf G}=&\{(0,0);(0,0);(0,0);(0,1);(0,0);(0,1);\cdots\}, 
\end{eqnarray}
%---------------------------------

\noindent where ${\bf C}=\Theta {\bf G}$, and the diffusive governing differential equation is:

%%EQUATION------------------------
\begin{eqnarray} 
\left( \frac{\partial}{\partial \Theta}-\nabla_{\bf r}^2\right)\Psi({\bf r},{\bf C})=0.
\label{GaussianDiffusionEqn}
\end{eqnarray}
%---------------------------------
\noindent where $\nabla_{\bf r}^2$ denotes the Laplacian in the x-y plane.

We now make a useful specialization which will be used for the remainder of this section on coherent fields, that only coherent aberrations are present in the linear shift-invariant imaging system. As previously mentioned, this amounts to specifying that the imaginary parts of all aberration coefficients vanish, thereby restricting ourselves to unitary diffraction operators.  

Under the above restriction, Eqs.~\ref{DMP03} and \ref{DMP09} assume the form of continuous aberration-coefficient output-field evolution published by Allen, Oxley and Paganin \cite{AllenOxleyPaganin}:

%%EQUATION------------------------
\begin{eqnarray} 
\left [\frac{\partial }{\partial \tau }-i\sum_{m,n}\left ( \frac{1}{i} \right )^{m+n}\frac{\partial C^{(R)}_{mn}}{\partial \tau }\frac{\partial ^{m}}{\partial x^{m}} \frac{\partial ^{n}}{\partial y^{n}} \right ]\Psi (\textbf{r},\textbf{C} )=0. \nonumber\\
\label{AWECSDaa5}
\end{eqnarray}
%---------------------------------

\noindent Here, the sum runs over all $C^{(R)}_{mn}$ for which $\partial C^{(R)}_{mn} / \partial \tau \neq 0$.

For a different set of expansion coefficients, Allen~$et$~$al$.~\cite{AllenOxleyPaganin} derived Eq.~\ref{AWECSDaa5} and showed that is reduces to the Schr$\ddot{\textup{o}}$dinger equation if time is considered as an aberration. It was also shown that Eq.~\ref{AWECSDaa5} leads to generalized forms of transport equation, such as the transport of intensity equation (TIE) of coherent paraxial scalar wave optics \cite{Madelung, TeagueTIEPaper}. For specimens that act as weak phase objects, Allen~$et$~$al$. \cite{AllenOxleyPaganin} derived a fourth-order transport equation for continuous variations of the spherical aberration, building upon earlier work of Lynch, Moodie and O'Keefe \cite{LynchMoodieOKeefe}.  

An important special case of Eq.~\ref{AWECSDaa5} corresponds to evolution along a particular direction $\tau\textbf{D}$ in aberration space, in which case $C^{(R)}_{mn}(\tau)=\tau \textup{D}_{mn}$ with $\textup{D}_{mn}$ independent of $\tau$. Equation \ref{AWECSDaa5} then reduces to:

%%EQUATION------------------------
\begin{eqnarray} 
 \left [  \frac{\partial }{\partial \tau }-i\sum_{m,n}\left ( \frac{1}{ i} \right )^{m+n}\textup{D}_{mn}\frac{\partial ^{m}}{\partial x^{m}} \frac{\partial ^{n}}{\partial y^{n}} \right ]\Psi (\textbf{r}, {\bf C})=0. \nonumber\\
\label{AWECSDaa6}
\end{eqnarray}
%---------------------------------

This form of the wave equation for evolution in aberration space is sufficiently general for us to derive the associated continuity and Hamilton--Jacobi equations. Multiplying through by $\Psi ^{*}$, writing $\Psi $ in polar form as $\Psi = \left |\Psi  \right | \exp (i \arg \Psi )$, and then separating real and imaginary parts, gives the aberration-space continuity equation \cite{AllenOxleyPaganin}:

%%EQUATION------------------------
\begin{eqnarray} 
\frac{\partial  }{\partial  \tau } \left |\Psi  \right |^{2}=-2\textup{Im}(\mathcal{M}) 
\label{AWECSDaa8}
\end{eqnarray}
%---------------------------------

\noindent and Hamilton--Jacobi (eikonal) equation:

%%EQUATION------------------------
\begin{eqnarray} 
\left |\Psi  \right |^{2}\frac{\partial  }{\partial  \tau }(\textup{arg}\Psi)=\textup{Re}(\mathcal{M}) 
\label{AWECSDaa9}
\end{eqnarray}
%---------------------------------

\noindent respectively. Here, 

%%EQUATION------------------------
\begin{eqnarray} 
\mathcal{M}=\Psi ^{*}\left [ \sum _{m,n} \left ( \frac{1}{i} \right )^{m+n} \textup{D}_{mn} \frac{\partial ^{m}}{\partial x^{m}} \frac{\partial ^{n}}{\partial y^{n}}\right ]\Psi .
\label{AWECSDaa10}
\end{eqnarray}
%---------------------------------

Alternatively, the term on the right-hand-side of Eq.~\ref{AWECSDaa8} can be written as the divergence of a current density vector $\textbf{J}$, enabling Eq.~\ref{AWECSDaa8} to be rewritten as:

%%EQUATION------------------------
\begin{eqnarray} 
\frac{\partial  }{\partial  \tau } \left |\Psi  \right |^{2}+\nabla_{\textbf{r}}\cdot \textbf{J} = 0. 
\label{ContinuityEquation}
\end{eqnarray}
%---------------------------------

\noindent In this case the current density is both spatially and aberration dependent and is given by 

%%EQUATION------------------------
\begin{eqnarray} 
\textbf{J}(\textbf{r} \mid \{\textup{D}_{mn}\}) &= &\sum _{m,n} \textup{Im}\left[ 2\Upsilon_{mn}  \Psi^{*} \mathcal{ \overrightarrow{D}}_{mn} \Psi \right] \nonumber\\
& &- \sum _{m,n} \textup{Im} \left [ \nabla_{\textbf{r}} \left ( \frac{ 2\Upsilon_{mn}}{\nabla_{\textbf{r}} ^{2}}  \left [\mathcal{ \overrightarrow{D}}_{mn} \Psi \cdot \nabla_{\textbf{r}}\Psi ^{*} \right ]  \right ) \right ] \nonumber\\
\label{CurrentDensityJ}
\end{eqnarray}
%---------------------------------

\noindent where $\Upsilon_{mn}=(1/i^{m+n}) \textup{D}_{mn}$. The vector operator $\mathcal{ \overrightarrow{D}}_{mn}$ is dependent on $\partial_{x}$ and $\partial_{y}$, these being shorthand for $\partial / \partial x$ and $\partial / \partial y$ with,

%%EQUATION------------------------
\begin{eqnarray} 
\mathcal{ \overrightarrow{D}}_{mn}(\partial_{x},\partial_{y})= \partial_{x}\left ( \frac{1}{\nabla_{\textbf{r}} ^{2}} \partial_{x}^{m} \partial_{y}^{n} \right ) \widehat{\textbf{x}} + \partial_{y} \left ( \frac{1}{\nabla_{\textbf{r}} ^{2}} \partial_{x}^{m} \partial_{y}^{n}\right ) \widehat{\textbf{y}}.  \nonumber\\
\label{CurrentOperator}
\end{eqnarray}
%--------------------------------- 
 
\noindent Here, $\widehat{\textbf{x}}$ and $\widehat{\textbf{y}}$ represent unit vectors associated with the Cartesian coordinates $\textbf{r}=(x,y)$. 

Once fully expanded, Eqs.~\ref{AWECSDaa8} and \ref{AWECSDaa9} are generally rather unwieldy, especially when written explicitly in terms of intensity $\textup{I}$ and phase $\varphi$ by writing the wave-function in polar form:  

%%EQUATION------------------------
\begin{eqnarray} 
\Psi (\textbf{r})= \sqrt{\textup{I}}\exp(i \varphi )
\label{AWECSDaa7}
\end{eqnarray}
%---------------------------------

\noindent where $\textup{I}= \left | \Psi \right |^{2}$, $\varphi =\arg{\Psi }$, and explicit functional dependence of $\Psi$, $\textup{I}$, $\varphi$ on $\textbf{r}$ and $C_{mn}$ has been dropped for clarity.

Note that the right side of the continuity equation (Eq.~\ref{ContinuityEquation}) will have a variety of transverse derivatives of the intensity, and transverse derivatives of the phase, of various orders that are directly related to the orders of the corresponding aberration coefficients.  This transport equation gives insight into the aberration-induced intensity contrast induced by the non-zero aberration coefficients. Under this view, the aberrated imaging system exhibits generalized differential phase contrast, with the $\tau$-rate-of-change of the output intensity depending on the transverse spatial derivatives, of various orders, of the phase of the field input into the system.  The same may be said of the more general form of the intensity transport equation, Eq.~\ref{AWECSDaa8}.

\subsection{Some special cases of the wave, continuity and Hamilton--Jacobi equations in aberration space}  \label{SubSec00C} 

Here we consider some special cases of Eqs. \ref{AWECSDaa6} through  \ref{AWECSDaa10}. This serves to exemplify the broad range of physical scenarios which these equations generalize.

\subsubsection{Paraxial scalar wave optics} \label{SubSecCase001}

\noindent For this case
  
%%EQUATION------------------------
\begin{eqnarray} 
\textbf{D}(\tau )\rightarrow \textbf{D}(z)=z \textbf{F},
\label{DP307}
\end{eqnarray}
%---------------------------------

\noindent for which Eq. \ref{AWECSDaa0} implies that the only non-vanishing aberration coefficients are:

%%EQUATION------------------------
\begin{eqnarray} 
\textup{D}_{02}=\textup{D}_{20}=-\frac{1}{2k}.
\label{DP308}
\end{eqnarray}
%---------------------------------

\noindent Hence Eq. \ref{AWECSDaa6} becomes:

%%EQUATION------------------------
\begin{eqnarray} 
\left ( 2ik\frac{\partial  }{\partial  z }+ \nabla_{\textbf{r}}^{2} \right ) \Psi =0,
\label{DP309}
\end{eqnarray}
%---------------------------------

\noindent which is the well-known parabolic equation of paraxial scalar wave optics \cite{SalehTeich}. This is the underpinning equation for Fresnel diffraction, to which Eq. \ref{AWECSD00} of the present paper provides a solution.

The associated continuity equation obtained by substituting Eqs. \ref{AWECSDaa10}, \ref{AWECSDaa7} and \ref{DP308} into Eq. \ref{AWECSDaa8}, is \citep{Madelung,TeagueTIEPaper}:

%%EQUATION-----------------------
\begin{eqnarray} 
- \nabla_{\textbf{r}}\cdot (\textup{I}\nabla_{\textbf{r}} \varphi )=k\frac{\partial \textup{I}}{\partial z}.
\label{DP3010}
\end{eqnarray}
%---------------------------------

Lastly, for the choice of non-zero aberration coefficients in Eq. \ref{DP308}, Eq. \ref{AWECSDaa9} becomes \cite{Gureyev1995}:

%%EQUATION-----------------------
\begin{eqnarray} 
\frac{\partial \varphi }{\partial z}=\frac{1}{2k}\left [ -\left | \nabla_{\textbf{r}} \varphi \right |^{2}+\frac{\nabla^{2}_{\textbf{r}} \sqrt{\textup{I}}}{\sqrt{\textup{I}}}  \right ], \: \textup{I} \neq 0. 
\label{DP3011}
\end{eqnarray}
%---------------------------------

This is the eikonal equation of paraxial geometric optics. In the short-wavelength limit, where the ``diffraction term'' $ {\nabla^{2}_{\textbf{r}}\sqrt{\textup{I}}}/{\sqrt{\textup{I}}}$ may be ignored, Eq. \ref{DP3011} is the paraxial approximation to 

%%EQUATION-----------------------
\begin{eqnarray} 
\left | \nabla_\textbf{R} \varphi \right |^{2}=k^{2},
\label{DP3012}
\end{eqnarray}
%---------------------------------

\noindent where $\nabla_\textbf{R}$ is the gradient in $\textbf{R}=(x,y,z)$. This equates to the statement that the spacing between adjacent wave crests, for gently deformed monochromatic plane waves, is approximately equal to the wavelength $\lambda$. The form given in Eq. \ref{DP3012} makes it clear that Eqs. \ref{AWECSDaa9} and \ref{DP3011} are indeed Hamilton--Jacobi equations: see e.g. $\S$10.8 of Goldstein's text \cite{GoldsteinBook}.

\subsubsection{Spherically-aberrated wave optics} \label{SubSecCase001}

Many optical imaging systems have some degree of spherical aberration present. This motivates us to write the wave equation specific to propagation in spherical aberration space, where

%%EQUATION-----------------------
\begin{eqnarray} 
\textbf{D}(\tau )\rightarrow \textbf{D}(C_{S})=C_{S} \textbf{S}.
\label{AWEASpheAbVector}
\end{eqnarray}
%---------------------------------

The differential equation, associated with this kind of propagation, is the following special case of Eq.~\ref{AWECSDaa6} \cite{AllenOxleyPaganin}:

%%EQUATION-----------------------
\begin{eqnarray} 
\left( 8ik^{3}\frac{\partial}{\partial C_S}-\nabla_{\bf r}^4\right)\Psi=0,
\end{eqnarray}
%---------------------------------

\noindent where $\nabla_{\bf r}^4=(\nabla_{\bf r}^2)^2 $. Equation~\ref{DP301aaa} becomes the convolution form of the diffraction integral associated with spherical aberration, Eq.~\ref{DMP00} becomes the real-space form of the propagator associated with spherical aberration, and Eqs.~\ref{DMP-3}--\ref{DMP-2} correspond to the operator form for diffraction in ``spherical aberration space''. The differential equation underpinning all of these diffraction integrals is of fourth order with respect to space, which may be compared to the fact that the parabolic equation associated with Fresnel diffraction (Eq.~\ref{DP309}) is of second order. 

The spherically-aberrated continuity and Hamilton--Jacobi equations respectively are:

%%EQUATION-----------------------
\begin{eqnarray} 
\frac{\partial \textup{I} }{\partial C_{S} }=-\frac{1}{4k^{3}}\nabla_{\textbf{r}}\cdot \textup{Im}\left [ \Psi^{*}\nabla_{\textbf{r}}\nabla^{2}_{\textbf{r}}\Psi + \textbf{g}\right ] \nonumber\\
 \textbf{g}=- \nabla_{\textbf{r}} \left [ \frac{1}{\nabla^{2}_{\textbf{r}}}\left [ \left ( \nabla_{\textbf{r}}\nabla^{2}_{\textbf{r}}\Psi \right ) \cdot \nabla_{\textbf{r}}\Psi^{*} \right ] \right ]% \nonumber\\
\label{SphericalContinuityEqn}
\end{eqnarray}
%---------------------------------

\noindent and,

%%EQUATION-----------------------
\begin{eqnarray} 
\textup{I} \frac{\partial \varphi  } {\partial C_{S } } =-\frac{1}{8k^{3}}\textup{Re} (\Psi^{*}\nabla^{4}_{\textbf{r} }\Psi).
\end{eqnarray}
%---------------------------------

For spherically-aberrated systems we make note of an interesting feature in Eq.~\ref{SphericalContinuityEqn}, namely the appearance of the vector $\textbf{g}$, which arises due to the second term of the current density $\textbf{J}$ (Eq.~\ref{CurrentDensityJ}). Under paraxial propagation (defocus) no vector of that nature appears in the continuity Eq.~\ref{DP3010} since the second term of $\textbf{J}$ vanishes. This $\textbf{g}$ term is somewhat non-intuitive, even for systems with rotationally symmetric aberrations. For rotationally symmetric aberrations that are higher than two in order (i.e. $m+n \geq 2$) $\textbf{g}$-like terms appear in the current density. 

\subsubsection{Non paraxial scalar wave optics} \label{SubSecCase002} 

\noindent For this case,

%%EQUATION-----------------------
\begin{eqnarray} 
\textbf{D}(\tau )\rightarrow \textbf{D}(z)=z \textbf{A}.
\label{DP3013}
\end{eqnarray}
%---------------------------------

\noindent We wish $\textbf{A}$ to correspond to non-paraxial scalar wave optics, under the assumption that all Fourier (plane-wave) components have a wave-vector $\textbf{k}_{\textbf{R}}=(k_{x},k_{y},k_{z})$ for which $\textup{Re }(k_{z})\geqslant 0$. This implies the wave to be forward propagating, but not necessarily paraxial. 

With a view to determining the components of $\textbf{A}$, with the ``A'' standing for ``angular-spectrum'' \cite{SalehTeich, MandelWolfBook}, we note the following binomial expansion \cite{AhmedKhan}:

%%EQUATION-----------------------
\begin{eqnarray} 
\sqrt{k^{2}-\left | \textbf{k}_{\textbf{r}} \right |^{2}}=k-\frac{\left | \textbf{k}_{\textbf{r}} \right |^{2}}{2k}-\frac{\left | \textbf{k}_{\textbf{r}} \right |^{4}}{8k^{3}}-\frac{\left | \textbf{k}_{\textbf{r}} \right |^{6}}{16k^{5}}-...
\label{DP3014}
\end{eqnarray}
%--------------------------------- 

Thus we can choose Eq. \ref{AWECSDaa6} to have the form: 

%%EQUATION-----------------------
\begin{eqnarray} 
\left ( \frac{\partial  }{\partial  z }-i\sqrt{k^{2}+\nabla_{\textbf{r}}^{2}}
 \right ) \Psi =0 
\label{DP3015}
\end{eqnarray}
%---------------------------------

\noindent if we take the components of $\textbf{A}$ as:

%%EQUATION-----------------------
\begin{eqnarray} 
A_{00}&=&k   \nonumber\\
A_{02}&=&A_{20}=\frac{-1}{2k} \nonumber\\
A_{40}&=&A_{04}=\frac{1}{2}A_{22}= \frac{-1}{8k^{3} } \nonumber\\
A_{60}&=&A_{06}=\frac{1}{3}A_{42}= \frac{1}{3}A_{24} = \frac{-1}{16k^{5}} \nonumber\\
 && \; \; \; \; \; \; \; \; \; \; \; \; \vdots 
\label{DP3016}
\end{eqnarray}
%--------------------------------- 

\noindent with all other components of $\textbf{A}$ being zero. 

The identification of $\textbf{A}$ with the ``aberration vector'', corresponding to non-paraxial scalar wave propagation for forward propagating coherent fields, then follows directly from either of the following arguments: (i) In vacuum, any forward-propagating plane wave $\exp\left [ i(k_{x}x+k_{y}y+k_{z}z) \right ]$ with $\textup{Re}(k_{z})\geqslant 0$ will obey Eq. \ref{DP3015}. The same plane wave obeys the Helmholtz equation of scalar wave optics, which is a linear differential equation and therefore obeys the superposition principle. Hence any solution to Eq. \ref{DP3015}, expressed as a superposition of the previously-mentioned forward-propagating plane waves, will solve the Helmholtz equation. Any boundary value of the field in the plane $z=0$ can be accommodated, provided the said boundary value meets the regularity conditions required for its Fourier transform to exist. Moreover, any forward propagating solution to the Helmholtz equation will obey Eq. \ref{DP3015}. (ii) Applying operator splitting to the Helmholtz equation    

%%EQUATION-----------------------
\begin{eqnarray} 
\left (\nabla_{\textbf{r}}^{2}+ \partial^{2} _{z}+k^{2} \right ) \Psi =0
\label{DP3017}
\end{eqnarray}
%--------------------------------- 

\noindent one obtains the formal factorization (see e.g. \cite{TeagueTIEPaper}) 

%%EQUATION-----------------------
\begin{eqnarray} 
\left ( \frac{\partial  }{\partial  z }+i\sqrt{k^{2}+\nabla_{\textbf{r}}^{2}}
 \right )\left ( \frac{\partial  }{\partial  z }-i\sqrt{k^{2}+\nabla_{\textbf{r}}^{2}}
 \right ) \Psi=0 . \nonumber\\
\label{DP3018}
\end{eqnarray}
%--------------------------------- 

\noindent We have two classes of solution:

%%EQUATION-----------------------
\begin{eqnarray} 
\left.\begin{matrix}
\left ( \frac{\partial  }{\partial  z }-i\sqrt{k^{2}+\nabla_{\textbf{r}}^{2}}
 \right )\Psi=0,\\ 
\left ( \frac{\partial  }{\partial  z }+i\sqrt{k^{2}+\nabla_{\textbf{r}}^{2}}
 \right ) \Psi=0,
\end{matrix}\right\}
\label{DP3019}
\end{eqnarray}
%--------------------------------- 

\noindent which respectively correspond to arbitrary forward-propagating and backward-propagating solutions to the Helmholtz equation.

It is interesting to note that, in passing from the paraxial ($\textup{\underline{F}resnel}$) ``aberration vector'' $\textbf{F}$ to the more general non-paraxial ($\textup{\underline{a}ngular}$ spectrum) ``aberration vector'' $\textbf{A}$, one has a process that is formally identical to adding higher-order aberrations to a paraxial imaging system. This is particularly evident in the decomposition:   

%%EQUATION-----------------------
\begin{eqnarray} 
\textbf{A}=\textbf{F}+\textbf{S}+\widetilde{\textbf{A}},
\label{DP3020}
\end{eqnarray}
%--------------------------------- 

\noindent where $\textbf{S}$ is the spherical-aberration ``aberration vector'' (see Eq. \ref{AWECSDaa2}) and the higher-order-terms ``aberration vector'' $\widetilde{\textbf{A}}$ is 

%%EQUATION-----------------------
\begin{eqnarray} 
\left.\begin{matrix}
\widetilde{A}_{00}=&k ,\\ 
\widetilde{A}_{mn}=&0\; \textup{if}\; 1\leq m+n\leq 5,  
\\ \widetilde{A}_{mn}=&A_{mn}\;\forall \; m+n \geq 6.
\end{matrix}\right\}
\label{DP3021}
\end{eqnarray}
%--------------------------------- 

The associated continuity equation, for non-paraxial forward propagating scalar waves, is given by the following special case of Eq. \ref{AWECSDaa8}:   

%%EQUATION-----------------------
\begin{eqnarray} 
\frac{\partial \textup{I}}{\partial z}=-2\textup{Im}\left ( \Psi^{*}\sqrt{k^{2}+\nabla_{\textbf{r}}^{2}}\Psi  \right ).  
\label{DP3022}
\end{eqnarray}
%--------------------------------- 

\noindent Here, $\sqrt{k^{2}+\nabla_{\textbf{r}}^{2}}$ is a pseudo-differential operator ultimately defined through its Fourier representation, namely $\sqrt{k^{2}+\nabla_{\textbf{r}}^{2}} \equiv \mathcal{F}^{-1} \sqrt{k^{2}-\left | \textbf{k}_{\textbf{r}} \right |^{2}} \mathcal{F}$.

We close this section by noting the following special case of the Hamilton--Jacobi equation Eq. \ref{AWECSDaa9}, for the non-paraxial fields studied in the present sub-section: 
 
%%EQUATION-----------------------
\begin{eqnarray} 
\textup{I}\frac{\partial \varphi }{\partial z}=\textup{Re}\left ( \Psi^{*}\sqrt{k^{2}+\nabla_{\textbf{r}}^{2}}\Psi  \right ).
\label{DP3026}
\end{eqnarray}
%--------------------------------- 

\subsection{Product form of the aberration propagator} \label{SubSec00D}

In the preceding sub-section, we noted that correction terms, due to transition from a paraxial to a non-paraxial formalism for forward propagating monochromatic scalar electromagnetic waves, are formally identical to taking a paraxial field and propagating it through an imaging system with specified aberrations given by the second and third terms on the right side of Eqn.~\ref{DP3020}. 

Indeed, Eq.~\ref{AWECSDa0} draws no distinction between (i) contributions from a given $C_{mn}$ which are due to the field equation obeyed by $\Psi$ in vacuum; (ii) contributions to a given aberration coefficient $C_{mn}$ that are due to the action of a shift-invariant imaging system. Each $C_{mn}$ can therefore be meaningfully decomposed as $C_{mn}=C^{(0)}_{mn}+C^{(1)}_{mn}$, where a zero superscript denotes the component of $C_{mn}$ due to point (i) above, and a unity superscript denotes the contribution due to point (ii) above.

The associated product form of the propagator in Eq. \ref{AWECSDa0}, namely        

%%EQUATION-----------------------
\begin{eqnarray} 
\exp\left [ i\sum_{m,n}C_{mn}k_{x}^{m}k_{y}^{n} \right ]&=
\exp\left [ i\sum_{m,n}C^{(0)}_{mn}k_{x}^{m}k_{y}^{n} \right ]  \nonumber\\
&\times \exp\left [ i\sum_{m,n}C^{(1)}_{mn}k_{x}^{m}k_{y}^{n} \right ], \nonumber\\
\label{DP3027}
\end{eqnarray}
%--------------------------------- 

\noindent may then be viewed as the Fourier representation of the Green's function for propagation of the free field, multiplied by the Fourier representation of the Green's function for passage through a linear shift-invariant system.   

This observation---that the field equations generate a set of fixed ``aberration coefficients'' $C^{(0)}_{mn}$ which may then be deformed via addition of a set of variable coefficients $C^{(1)}_{mn}$ if a shift-invariant imaging system is used to form images via the said field---is indicative of the generality of the formalism presented here.  A corollary is that the formalism of this paper may be applied to linear shift-invariant systems, and/or  vacuum propagation, for a variety of complex scalar fields obeying linear differential equations. Examples include paraxial scalar wave optics with or without an imaging system, non-paraxial scalar wave optics with or without an imaging system, forward propagating time-independent Klein--Gordon fields with or without an imaging system, diffusive fields with or without an imaging system, etc.

\section{Propagation of partially coherent fields in aberration space: A space--frequency description} \label{Sec01}

This section takes the formalism obtained from Eq.~\ref{AWECSDa0} and generalizes it to partially coherent complex scalar wavefields. We carry out this generalization by making use of the space-frequency description for partial coherence formulated by Wolf~\cite{WolfSpaceFreqPaper,WolfIntroToCoherenceTheory,Martinsson,MandelWolfBook}. Partially coherent wavefields are quantified in terms of the cross-spectral density (CSD). In physical terms, the CSD for a given angular frequency $\omega$ is obtained via an ensemble average over strictly monochromatic fields all having the same angular frequency $\omega$:

%%EQUATION-----------------------
\begin{eqnarray} 
W(\textbf{r}_{1},\textbf{r}_{2})=\left \langle \Psi^{*} (\textbf{r}_{1})\Psi (\textbf{r}_{2}) \right \rangle_{\omega }.
\label{AWECSD00a1}
\end{eqnarray}
%---------------------------------

\begin{figure}[h]
\centering
\includegraphics[scale=0.25]{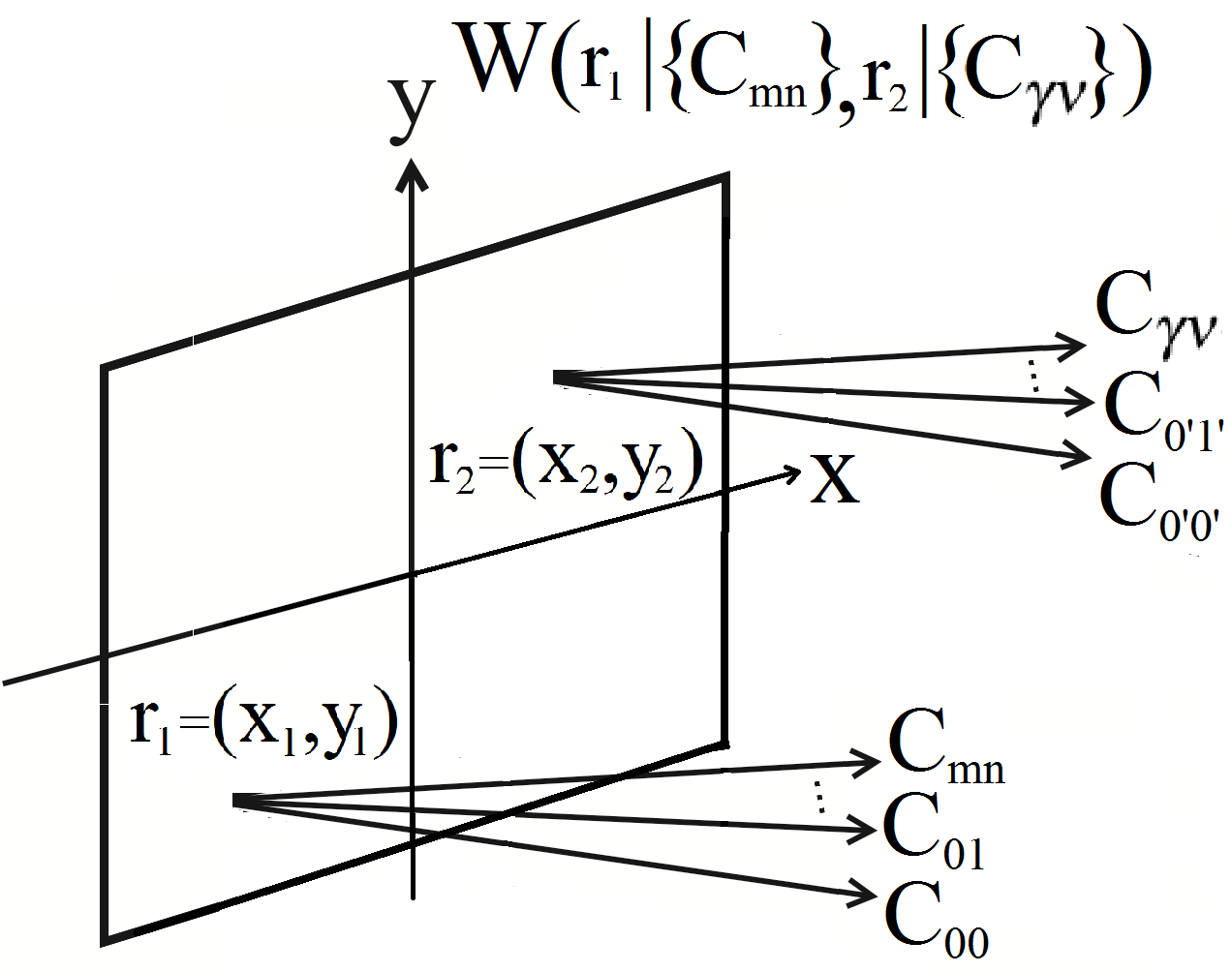}
\caption{ Illustration of 2-dimensional partially coherent waves propagating in aberration space. Note that the two sets of aberration variables, are distinguished by the presence or absence of primed indices.}\
\label{figure2}
\end{figure}

\noindent Here, the angular bracket $\left \langle  \right \rangle_{\omega}$ denotes the ensemble average. A key assumption of the formalism is that ensembles are statistically stationary and ergodic \cite{WolfSpaceFreqPaper}. From Eq.~\ref{AWECSDa0}, the forward scattering equation for the CSD transmitted by an arbitrary shift--invariant linear imaging system (see Fig.~\ref{figure2}) is: 

%%EQUATION-----------------------
\begin{eqnarray} 
&& W\left ( \textbf{r}_{1}\mid \left \{ C_{mn} \right \},\textbf{r}_{2}\mid \left \{ C_{\gamma\nu} \right \} \right ) =\frac{1}{(2\pi)^{2}}\iiiint_{-\infty}^{\infty}d\textbf{k}_{\textbf{r}_{1}}d\textbf{k}_{\textbf{r}_{2}} \nonumber\\
&&\:\:\:\:\:\:\:\: \times \exp \left [ i^{*}\sum_{m,n}C^{*}_{mn}k^{m}_{x_{1}}k^{n}_{y_{1}} +  i\sum_{\gamma, \nu}C_{\gamma\nu}k^{\gamma}_{x_{2}}k^{\nu}_{y_{2}}\right ] \nonumber\\
&&\:\:\:\:\:\:\:\: \times \exp[i^{*}\textbf{k}_{\textbf{r}_{1}}\cdot \textbf{r}_{1}+i\textbf{k}_{\textbf{r}_{2}}\cdot \textbf{r}_{2}] \nonumber\\
&&\:\:\:\:\:\:\:\: \times \left \langle \widehat{\Psi}^{*}( \textbf{k}_{\textbf{r}_{1}}\mid \left \{ C_{mn}\right \}=0) \widehat{\Psi}(\textbf{k}_{\textbf{r}_{2}}\mid \left \{ C_{\gamma\nu}\right \}=0 ) \right \rangle_{\omega}.
\label{AWECSD01}
\end{eqnarray}
%---------------------------------

\noindent Each spatial coordinate $\textbf{r}_{1}$ and $\textbf{r}_{2}$ is associated with individual sets of complex aberration coefficients $ \left \{ C_{mn} \right \} $ and $ \left \{ C_{\gamma\nu} \right \} $. $\textbf{k}_{\textbf{r}_{1}}$ and $\textbf{k}_{\textbf{r}_{2}}$ are Fourier space coordinates dual to $\textbf{r}_{1}$ and $\textbf{r}_{2}$. The scalar functions $\widehat{\Psi}^{*}( \textbf{k}_{\textbf{r}_{1}}\mid \left \{ C_{mn}\right \}=0)$ and  $\widehat{\Psi}(\textbf{k}_{\textbf{r}_{2}}\mid \left \{ C_{\gamma\nu}\right \}=0 )$ are Fourier transforms of the complex wavefields $\Psi( \textbf{r}_{1}\mid \left \{ C_{mn}\right \}=0)$ and $\Psi( \textbf{r}_{2}\mid \left \{ C_{\gamma\nu}\right \}=0)$.

%FIGURE--------------------------
%--------------------------------

\subsection{Partially coherent wave, continuity and Hamilton--Jacobi (eikonal) equations for propagation in aberration space} \label{SubSecPCA} 

The generalized wave equations for partially coherent fields in aberration space are obtained using a similar strategy to that used to derive Eqn~\ref{DMP03}. We apply the differential operators  $\partial / \partial C_{\gamma \nu}^{(R)}$, $\partial / \partial C_{\gamma \nu}^{(I)}$, $\partial / \partial C_{mn}^{(R)}$ and $\partial / \partial C_{mn}^{(I)}$ to both sides of Eq.~\ref{AWECSD01} which gives four separate sets of differential equations. We then invoke the chain-rule for two evolving parameters $\tau_{1}$ and $\tau_{2}$:

%%EQUATION-----------------------
\begin{subequations}
\begin{align}
\frac{\partial}{\partial\tau_{1}}W  = \sum_{m,n} \left[ 
\frac{\partial C_{mn}^{(R)}(\tau_{1})}{\partial\tau_{1}}\frac{\partial W}{\partial C_{mn}^{(R)}}
+\frac{\partial C_{mn}^{(I)}(\tau_{1})}{\partial\tau_{1}}\frac{\partial W}{\partial C_{mn}^{(I)}}\right] \label{ChainRulePartialCohCase:a}, \\
\frac{\partial}{\partial\tau_{2}}W  = \sum_{\gamma ,\nu } \left[ 
\frac{\partial C_{\gamma \nu }^{(R)}(\tau_{2})}{\partial\tau_{2}}\frac{\partial W}{\partial C_{\gamma \nu }^{(R)}}
+\frac{\partial C_{\gamma \nu }^{(I)}(\tau_{2})}{\partial\tau_{2}}\frac{\partial W}{\partial C_{\gamma \nu }^{(I)}}\right]. \label{ChainRulePartialCohCase:b}
\end{align}
\end{subequations}
%---------------------------------

\noindent The dependence on $\textbf{r}_{1}$, $\textbf{r}_{2}$, $\left \{ C_{\gamma\nu} \right \}$ and $\left \{C_{m n} \right \}$ has been dropped for simplicity. Once we substitute for $\partial W / \partial C_{mn}^{(R)}$, $\partial W / \partial C_{mn}^{(I)}$, $\partial W / \partial C_{\gamma \nu }^{(R)}$ and $\partial W / \partial C_{\gamma \nu }^{(I)}$ we arrive at a pair of generalized Wolf-type wave equations for the evolution of partially coherent wavefields in aberration space:

%%EQUATION-----------------------
\begin{subequations}
\begin{align}
\frac{\partial }{\partial \tau_{1} } W =-{\mathcal L} (\tau_{1})W  \label{AWECSD04:a}, \\
\frac{\partial }{\partial \tau_{2} } W={\mathcal L} (\tau_{2}) W , \label{AWECSD04:b}
\end{align}
\end{subequations}
%---------------------------------

\noindent where the operators ${\mathcal L}(\tau _{1})$ and ${\mathcal L}(\tau _{2})$ are:

%%EQUATION-----------------------
\begin{subequations}
\begin{align}
& {\mathcal L}(\tau _{1})=\sum_{m,n}\frac{1}{(i^{*})^{m+n}}\left [ i\frac{\partial C^{\textup{(R)}}_{mn}}{\partial \tau _{1}}+\frac{\partial C^{\textup{(I)}}_{mn}}{\partial \tau _{1}} \right ]\frac{\partial^{m} }{\partial x_{1}^{m} } \frac{\partial^{n} }{\partial y_{1}^{n} }  , \label{TempParCohOpp:a} \\
& {\mathcal L}(\tau _{2})=\sum_{\gamma,\nu}\frac{1}{i^{\gamma +\nu }}\left [ i\frac{\partial C^{\textup{(R)}}_{\gamma \nu}}{\partial \tau _{2}}-\frac{\partial C^{\textup{(I)}}_{\gamma \nu}}{\partial \tau _{2}} \right ]\frac{\partial^{\gamma} }{\partial x_{2}^{\gamma} } \frac{\partial^{\nu} }{\partial y_{2}^{\nu} }. \label{TempParCohOpp:b} 
\end{align}
\end{subequations}
%---------------------------------

Under the fully coherent limit the generalized Wolf-type Eqs.~\ref{AWECSD04:a} and \ref{AWECSD04:b} reduce to the less general Eq.~\ref{DMP03}. For the remainder of this section we will restrict ourselves to linear shift-invariant imaging systems where only coherent aberrations are present (i.e. $C_{\gamma \nu }=C^{(R)}_{\gamma \nu }$, $C_{mn}= C^{(R)}_{mn }$ ) and incoherent aberrations vanish (i.e. $C^{(I)}_{\gamma \nu }=0$, $C^{(I)}_{mn}=0$ ). Also, we consider sets of aberration coefficients that evolve continuously as a function of the real non-negative parameters $\tau_{1}$ and $\tau_{2}$. That is

%%EQUATION-----------------------
\begin{eqnarray} 
\textbf{D}(\tau_{1})= \left \{  C^{(R)}_{mn }(\tau_{1}) \right \},\; \textbf{D}(\tau_{2})= \left \{ C^{(R)}_{\gamma \nu }(\tau_{2}) \right \} .   
\label{AWECSCaa01}
\end{eqnarray}
%---------------------------------

Under such restrictions Eqs.~\ref{AWECSD04:a} and \ref{AWECSD04:b} become:

%%EQUATION-----------------------
\begin{subequations}
\begin{align}
& \left [ \frac{\partial}{\partial \tau_{1} } +i\sum_{m,n}\left ( \frac{1}{i^{*}} \right )^{m+n}\frac{\partial C^{(R)}_{mn }}{\partial \tau_{1} } \frac{\partial^{m} }{\partial x_{1}^{m} } \frac{\partial^{n} }{\partial y_{1}^{n} }  \right ]W=0    \label{AWECSC:a}, \\
& \left [ \frac{\partial}{\partial \tau_{2} } -i\sum_{\gamma ,\nu }\left ( \frac{1}{i} \right )^{\gamma+\nu}\frac{\partial C^{(R)}_{\gamma \nu}}{\partial \tau_{2} }\frac{\partial^{\gamma} }{\partial x_{2}^{\gamma} } \frac{\partial^{\nu} }{\partial y_{2}^{\nu} } \right ]W=0. \label{AWECSC:b}
\end{align}
\end{subequations}
%---------------------------------

\noindent The above expressions are a generalized version of the result derived in Eq.~\ref{AWECSDaa5} by Allen~$et$~$al$.~\cite{AllenOxleyPaganin} where partial coherence is ignored. 

For the case where the parameters $\tau_{1}$ and $\tau_{2}$ evolve along a specific direction, that is,

%%EQUATION-----------------------
\begin{eqnarray} 
 C^{(R)}_{m n}(\tau_{1})=\tau_{1}\textup{D}_{mn} ,\;   C^{(R) }_{\gamma \nu }(\tau_{2})=\tau_{2}\textup{D}_{\gamma \nu },
\label{AWECSCaa02}
\end{eqnarray}
%---------------------------------

\noindent Eqs.~\ref{AWECSC:a} and \ref{AWECSC:b} reduce to  
 
%%EQUATION-----------------------
\begin{subequations}
\begin{align}
& \left [\frac{\partial}{\partial \tau_{1} }+i\sum_{m,n}\left ( \frac{1}{i^{*}} \right )^{m+n}\textup{D}_{mn}  \frac{\partial^{m} }{\partial x_{1}^{m} } \frac{\partial^{n} }{\partial y_{1}^{n} } \right ]W=0    \label{AWECSC2:a}, \\
& \left [ \frac{\partial}{\partial \tau_{2} }-i\sum_{\gamma ,\nu }\left ( \frac{1}{i} \right )^{\gamma+\nu}\textup{D}_{\gamma \nu} \frac{\partial^{\gamma} }{\partial x_{2}^{\gamma} } \frac{\partial^{\nu} }{\partial y_{2}^{\nu} } \right ]W=0. \label{AWECSC2:b}
\end{align}
\end{subequations}
%---------------------------------

The associated continuity and Hamilton--Jacobi equations for partially coherent fields evolving in aberration space can be obtained by multiplying Eqs.~\ref{AWECSC2:a} and \ref{AWECSC2:b} by $W ^{*}$, followed by separating real and imaginary parts. This gives the continuity equations

%%EQUATION-----------------------
\begin{subequations}
\begin{align}
& \frac{\partial  \left | W \right |^{2} }{\partial \tau_{1} } =2 \textup{Im} (\mathcal{M}_{1})    \label{AWECSC3:a}, \\
& \frac{\partial  \left | W \right |^{2} }{\partial \tau_{2} } =-2 \textup{Im} (\mathcal{M}_{2}) \label{AWECSC3:b},
\end{align}
\end{subequations}
%---------------------------------

\noindent and the Hamilton-Jacobi (eikonal) equations

%%EQUATION-----------------------
\begin{subequations}
\begin{align}
& \left | W \right |^{2} \frac{\partial }{\partial \tau_{1} } (\textup{arg} W )  =- \textup{Re} (\mathcal{M}_{1})   \label{AWECSC4:a}, \\
& \left | W \right |^{2}  \frac{\partial }{\partial \tau_{2} }(\textup{arg} W )  = \textup{Re} (\mathcal{M}_{2})\label{AWECSC4:b},
\end{align}
\end{subequations}
%---------------------------------

\noindent with, 

%%EQUATION-----------------------
\begin{subequations}
\begin{align}
& \mathcal{M}_{1}=W^{*}\left [ \sum_{m,n}\left ( \frac{1}{i^{*}} \right )^{m+n}\textup{D}_{mn}\frac{\partial^{m} }{\partial x_{1}^{m} } \frac{\partial^{n} }{\partial y_{1}^{n} }  \right ]W \label{AWECSC5:a}, \\
& \mathcal{M}_{2}=W^{*}\left [ \sum_{\gamma ,\nu }\left ( \frac{1}{i} \right )^{\gamma+\nu}\textup{D}_{\gamma \nu}\frac{\partial^{\gamma} }{\partial x_{2}^{\gamma} } \frac{\partial^{\nu} }{\partial y_{2}^{\nu} } \right ]W. \label{AWECSC5:b}
\end{align}
\end{subequations}
%---------------------------------

Note that Eqs.~\ref{AWECSC3:a} and \ref{AWECSC3:b} can be rewritten as a divergence of a current density vector field similarly to Eq.~\ref{ContinuityEquation}, so that expressions for such currents can be derived. These will not be given in this manuscript due to lengthy computation required. An exception is made for the simple case of free-space propagation in Sec.~\ref{SubSecParCase001} below as the result is compact. Similar manipulations to that used in deriving Eq.~\ref{CurrentDensityJ} can be applied to derive such general current density expressions.

From this point onwards we examine particular cases of aberration as was done in Sec. \ref{SubSec00C} for fully coherent fields. This following section mirrors the special cases in Sec.~\ref{SubSec00C}, however for partially coherent fields rather than fully coherent fields.       

\subsection{Some special cases of the partially coherent wave, continuity and Hamilton--Jacobi equations in aberration space}  \label{SubSecPCB} 

We consider some special cases of Eqs.~64-67 to demonstrate the broad range of physical scenarios to which the present formalism may be applied.

\subsubsection{Partially coherent paraxial scalar wave optics} \label{SubSecParCase001}

For the specific case where only defocus aberration is present, that is,

%%EQUATION-----------------------
\begin{eqnarray} 
\textbf{D}(\tau_{1} )\rightarrow \textbf{D}(z_{1})=z_{1} \textbf{F},\nonumber\\
\textbf{D}(\tau_{2} )\rightarrow \textbf{D}(z_{2})=z_{2} \textbf{F},
\label{AWECSD05}
\end{eqnarray}
%---------------------------------

\noindent Eqs. \ref{AWECSC2:a} and \ref{AWECSC2:b} reduce to the well-known paraxial Wolf equations \cite{Petrucelli2013}:

%%EQUATION-----------------------
\begin{subequations}
\begin{align}
& \left ( 2ik\frac{\partial}{\partial z_{1}}-\nabla^{2}_{\textbf{r}_{1}}  \right ) W=0  \label{AWECSD07:a}, \\
& \left ( 2ik\frac{\partial}{\partial z_{2}} +\nabla^{2}_{\textbf{r}_{2}} \right )W=0. \label{AWECSD07:b}
\end{align}
\end{subequations}
%---------------------------------

\noindent The associated continuity Eqs.~\ref{AWECSC3:a} and \ref{AWECSC3:b} become,

%%EQUATION-----------------------
\begin{subequations}
\begin{align}
& \frac{\partial  \left | W \right |^{2}}{\partial z_{1}}= \frac{1}{k}\nabla_{\textbf{r}_{1}}\cdot  \textup{Im}\left ( W^{*} \nabla_{\textbf{r}_{1}} W\right )   \label{AWEAS0003:a}, \\
& \frac{\partial  \left | W \right |^{2}}{\partial z_{2}}= -\frac{1}{k}\nabla_{\textbf{r}_{2}}\cdot  \textup{Im}\left ( W^{*} \nabla_{\textbf{r}_{2}} W\right ). \label{AWEAS0003:b}
\end{align}
\end{subequations}
%---------------------------------

\noindent Here, $\textup{Im}\left ( W^{*} \nabla_{\textbf{r}_{1}} W\right )$ and $\textup{Im}\left ( W^{*} \nabla_{\textbf{r}_{2}} W\right )$ are directly related to the coherence current densities $\textbf{J}$ for scalar partially coherent forward paraxial wave propagation \cite{BerryOpticalCurrents,WangTakedaPRL}. 

The paraxial partially coherent Hamilton--Jacobi (eikonal) Eqs~\ref{AWECSC4:a} and \ref{AWECSC4:b} reduce to:      

%%EQUATION-----------------------
\begin{subequations}
\begin{align}
& \frac{\partial \left ( \textup{arg}W \right )}{\partial z_{1}}  = -\frac{1}{2k}\left [ 
\frac{ \nabla_{\textbf{r}_{1}}^{2}\left | W \right |}{\left | W \right |}-\left |  \nabla_{\textbf{r}_{1}} \left ( \textup{arg}W \right )\right |^{2} \right ]   \label{AWEAS0004:a}, \\
& \frac{\partial \left ( \textup{arg}W \right )}{\partial z_{2}} = \frac{1}{2k}\left [ 
\frac{ \nabla_{\textbf{r}_{2}}^{2}\left | W \right |}{\left | W \right |}-\left |  \nabla_{\textbf{r}_{2}} \left ( \textup{arg}W \right )\right |^{2} \right ].   \label{AWEAS0004:b}
\end{align}
\end{subequations}
%---------------------------------

\subsubsection{Partially coherent spherically-aberrated wave optics} \label{SubSecParCase002}

The pair of Wolf-type equations specific to spherical aberration $C_{S}$ implies the following case:

%%EQUATION-----------------------
\begin{eqnarray} 
\textbf{D}(\tau_{1} )\rightarrow \textbf{D}(C_{S_{1}})=C_{S_{1}} \textbf{S} ,\nonumber\\
\textbf{D}(\tau_{2} )\rightarrow \textbf{D}(C_{S_{2}})=C_{S_{2}} \textbf{S}.
\label{AWEAS0001}
\end{eqnarray}
%---------------------------------

\noindent Here, Eqs.~\ref{AWECSC2:a} and \ref{AWECSC2:b} reduce to the following pair of fourth order differential equations:  

%%EQUATION-----------------------
\begin{subequations}
\begin{align}
& \left ( 8ik^{3}\frac{\partial}{\partial C_{S_{1}} }+\nabla^{4}_{\textbf{r}_{1}} \right )W=0      \label{AWECSD08:a}, \\
& \left ( 8ik^{3}\frac{\partial}{\partial C_{S_{2}} }-\nabla^{4}_{\textbf{r}_{2}} \right )W=0,     \label{AWECSD08:b}
\end{align}
\end{subequations}
%---------------------------------

\noindent with the associated continuity equations being,

%%EQUATION-----------------------
\begin{subequations}
\begin{align}
& \frac{\partial \left | W \right |^{2} }{\partial C_{S_{1}} } =\frac{1}{4k^{3}} \textup{Im} (W^{*}\nabla^{4}_{\textbf{r}_{1}}W)  \label{AWECSD08CE:a}, \\
& \frac{\partial \left | W \right |^{2} }{\partial C_{S_{2}} }=-\frac{1}{4k^{3}}\textup{Im} (W^{*}\nabla^{4}_{\textbf{r}_{2}}W), \label{AWECSD08CE:b}
\end{align}
\end{subequations}
%---------------------------------

\noindent and the corresponding Hamilton--Jacobi (eikonal) equations:
  
%%EQUATION-----------------------
\begin{subequations}
\begin{align}
& \left | W \right |^{2}\frac{\partial (\textup{arg} W ) }{\partial C_{S_{1}} }  = \frac{1}{8k^{3}}\textup{Re} (W^{*}\nabla^{4}_{\textbf{r}_{1}}W)  \ \label{AWECSD08Eik:a}, \\
& \left | W \right |^{2} \frac{\partial (\textup{arg} W ) }{\partial C_{S_{2}} } =-\frac{1}{8k^{3}}\textup{Re} (W^{*}\nabla^{4}_{\textbf{r}_{2}}W). \label{AWECSD08Eik:b}
\end{align}
\end{subequations}
%---------------------------------

\subsubsection{Partially coherent non-paraxial scalar wave optics} \label{SubSecParCase002}

Finally, we revisit the case where the fields are forward free-space propagating but not necessarily paraxial. Here, we have the set of infinitely many rotationally symmetric aberrations given by

%%EQUATION-----------------------
\begin{eqnarray} 
\textbf{D}(\tau_{1} )\rightarrow \textbf{D}(z_{1})=z_{1} \textbf{A}  ,\nonumber\\
 \textbf{D}(\tau_{2} )\rightarrow \textbf{D}(z_{2})=z_{2}\textbf{A} . 
\label{AWEAS0002}
\end{eqnarray}
%---------------------------------

Equations \ref{AWECSC2:a} and \ref{AWECSC2:b} reduce to

%%EQUATION-----------------------
\begin{subequations}
\begin{align}
& \left (\frac{\partial }{\partial z_{1} }+i\sqrt{k^{2}+\nabla_{\textbf{r}_{1}}^{2}}
 \right )W=0  , \label{AWECSD00W:a} \\
& \left (\frac{\partial }{\partial z_{2} }-i\sqrt{k^{2}+\nabla_{\textbf{r}_{2}}^{2}}
 \right )W=0, \label{AWECSD00W:b}
\end{align}
\end{subequations}
%---------------------------------

\noindent which are forward-propagating solutions to the Wolf equations:

%%EQUATION-----------------------
\begin{subequations}
\begin{align}
& \left ( k^{2}+\nabla_{\textbf{R}_{1}}^{2}\right )W=0 , \label{AWECSD00W:a1} \\
&  \left ( k^{2}+\nabla_{\textbf{R}_{2}}^{2}\right )W=0. \label{AWECSD00W:b1}
\end{align}
\end{subequations}
%---------------------------------

\noindent where $\nabla_{\textbf{R}}^{2}= ( \partial / \partial x)^{2}+( \partial / \partial y)^{2}+( \partial / \partial z)^{2}$,

The continuity equations for the non-paraxial case become:

%%EQUATION-----------------------
\begin{subequations}
\begin{align}
& \frac{\partial \left | W \right |^{2}}{\partial z_{1} } =2 \textup{Im}\left ( W^{*}\sqrt{k^{2}+\nabla_{\textbf{r}_{1}}^{2}}W \right ),\label{AWECSD00WCE:a} \\
& \frac{\partial \left | W \right |^{2}}{\partial z_{2} } =-2 \textup{Im}\left ( W^{*}\sqrt{k^{2}+\nabla_{\textbf{r}_{2}}^{2}}W \right ). \label{AWECSD00WCE:b}
\end{align}
\end{subequations}
%---------------------------------

\noindent Similarly, the Hamilton--Jacobi equations are:

%%EQUATION-----------------------
\begin{subequations}
\begin{align}
& \left | W \right |^{2}\frac{\partial \left ( \textup{arg}W \right )}{\partial z_{1} }   =\textup{Re}\left ( W^{*}\sqrt{k^{2}+\nabla_{\textbf{r}_{1}}^{2}}W \right )  , \label{AWECSD00WEIK:a} \\
& \left | W \right |^{2}\frac{\partial \left ( \textup{arg}W \right )}{\partial z_{2} }   =-\textup{Re}\left ( W^{*}\sqrt{k^{2}+\nabla_{\textbf{r}_{2}}^{2}}W \right ). \label{AWECSD00EIK:b}
\end{align}
\end{subequations}
%---------------------------------

\section{Discussion and Summary} \label{Sec03}

This study has revisited propagation of fully coherent complex scalar wave-fields through linear but otherwise arbitrary shift-invariant optical imaging systems. This constitutes a generalized diffraction integral (Eq.~\ref{AWECSDa0}) for propagating the wave-functions into an infinite-dimensional aberration space with coordinates given by values defined by the set of aberration coefficients $\left \{C_{m n} \right \} \equiv {\bf C}$, each of which describes a given state of the linear but otherwise arbitrary optical imaging system. This generalized diffraction integral was seen to imply an infinite number of differential equations which govern the evolution and energy transportation of wavefields along each ``direction'' in aberration space (Eq.~\ref{DMP03}).  A distinction was drawn between the infinite multiplicity of vacuum field equations permitted by this formalism (e.g. Helmholtz, paraxial equation, etc.) and the infinite multiplicity of aberrated shift-invariant imaging systems through which such fields may propagate (e.g. systems with defocus, spherical aberration, Gaussian blur etc.). Transport, Hamilton--Jacobi and continuity equations also were derived from this generalized diffraction integral (Eqs.~\ref{AWECSDaa8}, \ref{AWECSDaa9}, and \ref{ContinuityEquation} respectively).  

We saw that every aberration $C_{mn}$ associated with our formalism may be written as sum of a coefficient $C_{mn}^{(0)}$ contributing to the underpinning vacuum field equation, and a coefficient $C_{mn}^{(1)}$ corresponding to any shift invariant imaging system designed to form images using such radiation (cf. Eq. \ref{DP3027}). These imaging systems were not restricted by the requirement for unitarity. Stated differently, these imaging systems were not restricted to coherent aberrations alone, although we did require the system to be non-amplifying in order for non-physical divergences to be avoided (Eq.~\ref{AbbSpaceIsAHalfSpace}). Bearing the preceding comments in mind, the present paper may be viewed as a step towards a generalized wave theory for propagating arbitrary complex coherent scalar fields through arbitrary shift-invariant systems.  

This generalization was itself generalized, in passing from coherent fields to the two-point correlation function (cross-spectral density) used in the present paper to describe statistically stationary partially coherent fields streaming through aberrated shift-invariant imaging systems.  The second part of this investigation studied the propagation of such two-point correlation functions, associated with the space--frequency description of partially coherent complex scalar optical wavefields evolving into a space of infinitely many aberrations. For this case, a two-point-correlation-function diffraction integral was developed, where each pair of points ${\bf{r}}_1$ and ${\bf{r}}_2$ has an associated infinite set of aberrations, denoted $\{C_{\gamma\nu}\}$ and $\{C_{mn}\}$, respectively (Eq.~\ref{AWECSD01}). This lead us to an infinite number of Wolf-type wave--diffusion differential equations (Eqs.~\ref{AWECSD04:a} and \ref{AWECSD04:b}) together with their associated continuity and Hamilton--Jacobi equations (Eqs. \ref{AWECSC3:a}, \ref{AWECSC3:b}, \ref{AWECSC4:a}, \ref{AWECSC4:b}, respectively).  The wave-like Wolf-type equations were associated with coherent aberrations and contained the well-known form of the Wolf equations as a special case (Eqs.~\ref{AWECSD00W:a1} and \ref{AWECSD00W:b1}), while the diffusive Wolf-type equations were seen to be associated with incoherent aberrations. We demonstrated that for a system containing specific aberration types, the formalism reduces to well-established field equations such as the pair of paraxial, non-paraxial and spherically aberrated Wolf equations \citep{MandelWolfBook,Petrucelli2013} -- see Eqs.~\ref{AWECSD07:a}, \ref{AWECSD07:b}, \ref{AWECSD00W:a1}, \ref{AWECSD00W:b1}, and  \ref{AWECSD08:a}, \ref{AWECSD08:b}, respectively.

In addition to being applicable to the forward problem of arbitrary shift-invariant linear imaging systems streaming coherent or partially coherent scalar fields governed by arbitrary vacuum field equations, our formalism may also be applicable to the associated inverse problem of phase retrieval. With this in mind, we recall the fact that the transport/continuity equation associated with paraxial coherent fields, namely the transport-of-intensity equation \cite{TeagueTIEPaper} (see Eq.~\ref{DP3010} above), has been used as the starting point for a variety of phase retrieval schemes \cite{PaganinXRayBook,TeagueTIEPaper,Gureyev1995,PaganinNungent,BartyNungent}. With reference to Eq.~\ref{DP3010}, the key idea is that both the intensity and the derivative with respect to defocus of the intensity are physically measurable quantities, while at optical and higher temporal frequencies the wavefield oscillations are too rapid for the phase to be directly measurable. One can then treat the phase of the input field in Eq.~\ref{DP3010} as an unknown, with this elliptic second-order partial differential equation to be solved for the said phase map subject to both the measured intensity data and suitable boundary conditions \citep{TeagueTIEPaper,Gureyev1995}. There may be some scope for such a phase-retrieval scenario to be generalized to the case of phase retrieval using arbitrary shift-invariant aberrated imaging systems. Specifically, one could use the generalized intensity-transport equation~\ref{AWECSDaa8}  as a starting point for the inverse problem of phase retrieval. Under this view, the left side of Eq.~\ref{AWECSDaa8}  can be measured directly; Eq.~\ref{AWECSDaa7} could be substituted into Eq.~\ref{AWECSDaa8} via Eq.~\ref{AWECSDaa10}; provided that the input intensity could be directly measured, the only remaining unknown in the resulting partial differential equation will be the unknown phase. Moreover, if, for example, it were known {\em a priori} that the object at the entrance surface of the imaging system was thin, entirely contained within the field of view of the system, and illuminated with uniform intensity normally-incident plane waves, and that the aberrations were not so strong as to be in a holographic regime, then zero boundary conditions (trivial Dirichlet boundary conditions) could be assumed for the resulting partial differential equation, in which the output phase would be the only unknown. With the output phase thus inferred, and the output intensity measured, the inverse system propagator could be used to infer the phase of the input field. While the investigation of such phase-retrieval scenarios in the context of arbitrary aberrated shift-invariant imaging systems is beyond the scope of the present paper, it would form an interesting topic for further investigation, which would be considerably more general than the initial steps in this direction due to Paganin and Gureyev \cite{AbbBalancingPaganinGureyev}.

We have focused on the cross-spectral density as a descriptor of the two-point correlation properties associated with statistically stationary partially coherent fields. This statement invites at least five obvious avenues for further investigation. (i) For the case of Gaussian statistics, the Gaussian moment theorem (Siegert relation) \cite{GoodmanBook} implies that all higher-order correlation functions either vanish, or may be determined as a function of the corresponding two-point correlation functions. For non-Gaussian statistics, it would be interesting to investigate higher-order correlation functions, such as those of fourth order, which are of importance in contexts such as the quantum optics of non-classical light fields \cite{MandelWolfBook}. For fields obeying Gaussian statistics, using the Siegert relation to construct fourth-order field correlations associated with intensity--intensity correlations may be useful in generalizations of both the Hanbury Brown--Twiss effect and ghost imaging (see e.g. the review by Shirai \cite{Shirai}, together with primary references therein). (ii) The assumption of statistical stationarity, as employed for all partially coherent fields considered in the present paper, could be dropped in favour of less restrictive assumptions such as cyclo-stationarity \cite{Gardner}. (iii) Scalar fields could be replaced with multi component fields such as tensor or spinor fields, thereby extending our formalism into one capable of considering e.g. the effects of partial polarization \citep{WolfIntroToCoherenceTheory} for electromagnetic and other tensorial fields, and the effects of partial polarization for electron and other spinorial fields. (iv) Cross-spectral densities are not the only means for visualizing two-point correlation properties associated with statistically stationary scalar fields. Such coherence functions can be readily transformed into several different coherence functions, such as the mutual coherence function \cite{BornWolf1999}, the Wigner function \cite{Wigner}, the ambiguity function \cite{Papoulis}, the generalized radiance \citep{Walther,MarchandWolf} etc. We refer the reader to the excellent review by Alonso \cite{Alonso}, which considers these and other two-point coherence functions from a single cohesive perspective.  It would be interesting to construct and study such additional representations of the two-point correlation properties of statistically stationary fields, in the setting of their evolution upon passage through aberrated shift-invariant linear imaging systems, for the additional insights that such alternative representations may provide. (v) The generalized Wolf-type equations could be readily transformed from their Helmholtz-like space--frequency-domain form to a d'Alembert-like space--time form.  This could be done for both the scalar correlation function considered in the present paper, and for the previously mentioned generalization to multi-component fields and their associated tensorial cross-spectral densities.

Returning to the wavefunctions (for coherent fields) and cross-spectral densities (for partially coherent fields) as studied in the present paper, it would be interesting to investigate both from the perspectives of singular optics \cite{Nye}. (i) In the coherent case of our formalism, for all but the most trivial fields one would expect the aberration space to often be permeated with nodal lines of the wavefunction $\Psi$, with such nodal lines of vanishing intensity threading multi-valued phase-vortex cores associated with screw-type dislocations in $\textrm{arg} \Psi$ \citep{Nye,Dirac,Simula,PaganinXRayBook}. In the corresponding partially coherent case of our formalism, one would have analogous screw-type dislocations---known as coherence vortices \citep{SchoutenWolf,GburVisser}---associated with the phase of the cross-spectral density, threaded by nodal lines of vanishing two-point correlation \cite{MarashinghePaganin}. The topological dynamics of such nodal-line networks, which would trace out high-dimensional zero sheets in aberration space, would be an interesting avenue for future work. (ii) The previously mentioned phase dislocations are singularities of the wave theory which vanish in the geometrical optics limit where the wavelength of the field may be assumed negligibly small \cite{Nye}. In such a limit, which may also be associated with coarse-graining over a sufficiently large spatial extent for explicitly wave-optical effects to be smeared out, a new and complementary form of singularity emerges, namely the caustic \cite{Nye}. These are envelopes of ray families, along which the intensity diverges, that may be categorized into a number of equivalence classes using the machinery of catastrophe theory \cite{Nye,MBerry}. One would expect analogous structures to emerge in the infinite-dimensional aberration space utilized in the present paper, in the geometric-optics limit for coherent fields. Interestingly, such a stage would support an exotic hierarchy of caustics (optical catastrophes) of arbitrarily high dimensionality. Whether analogous caustical structures exist in the partially coherent case is unclear to the present authors. As with the previously mentioned nodal-line networks, the caustic networks considered here would also be expected to exhibit rich topological dynamics as they evolve through aberration space. 

We close with three miscellaneous remarks: 

Remark \#1: It would be interesting to undertake an asymptotic analysis of the generalized cross-spectral diffraction integral given in Eq.~\ref{AWECSD01}, for special cases of the input cross-spectral density such as that furnished by the Schell model \cite{MandelWolfBook}. One could also assume the input disturbance to be delta correlated, in such a setting, thereby leading to a generalized form of the van Cittert--Zernike theorem \citep{Cittert,Zernike1938}. 

Remark \#2: Passage from pure defocus for coherent fields ($C_{20}=C_{02}=-z/2k$), to pure Gaussian blur ($C_{20}=C_{02}=i\Theta$), was seen to convert a parabolic (Schr\"odinger) equation (Eq.~\ref{DP309}) into a diffusion equation (Eq.~\ref{GaussianDiffusionEqn}). Clearly, this is directly related to the concept of a Wick rotation in quantum field theory \cite{Maggiore}, a connection which may warrant further investigation. 

Remark \#3: Returning attention again to Eq.~\ref{AWECSD01} for the propagation of two-point field correlations into aberration space, it was clear from this equation's derivation, that two sets of aberration coefficients parameterize this diffraction integral. Given that (i) each set of aberrations can be viewed as a generalization of the defocus aberration associated with free-space propagation of paraxial monochromatic fields obeying the parabolic equation (Eq.~\ref{DP309}), and given that (ii) the free space propagation of the cross-spectral density $W$ will evolve $W({\bf r}_1,{\bf r}_2)$ to $W({\bf r}_1,{\bf r}_2; z_1, z_2)$, where $z_1$ and $z_2$ respectively denote the propagation distances (defoci) for the first and second spatial coordinates in the unpropagated/input two-point correlation function $W({\bf r}_1,{\bf r}_2)$, it is natural that (iii) the generalization given by Eq.~\ref{AWECSD01} should replace $z_1$ with one particular set of aberration coefficients $\{C_{mn}\}\equiv {\bf C}_1$, and $z_2$ with an in-general different set of aberration coefficients $\{C_{\gamma\nu}\}\equiv {\bf C}_2$. If one considers the special case where the set ${\bf C}_1$ is constrained to be equal to the set ${\bf C}_2$, then one is restricting consideration to both points of the propagated spectral density being constrained to lie over the exit surface of the same (single) aberrated shift-invariant imaging system, albeit with such a system being in an arbitrary state. A means of physically realizing the more general case of Eq.~\ref{AWECSD01} where ${\bf C}_1 \ne {\bf C}_2$, albeit subject to the additional restriction that at least one of the aberration vectors (say, ${\bf C}_1$) contains no incoherent aberrations, would be for imaging systems characterized by ${\bf C}_1$ and ${\bf C}_2-{\bf C}_1$ respectively to be placed in series, with the point ${\bf r}_1$ being at the exit surface of the first system, and the point ${\bf r}_2$ being at the exit surface of the second system.

\end{document}